%% file: Paper.tex
\documentclass[twocolumn,resetfootnote]{aastex7}

\usepackage{hyperref}

\usepackage{siunitx}
\DeclareSIUnit \parsec {pc}
\DeclareSIUnit \arcminSpelled {arcmin}
\DeclareSIUnit \arcsecSpelled {arcsec}
\DeclareSIUnit \erg {erg}

\shorttitle{Separating Light of AGN and Hosts}
\shortauthors{Dewsnap et al.}

\graphicspath{{./}{Figures/}}

\begin{document}

\title{Separating the Optical to Near-Infrared Light of AGN and Their Host Galaxies}

\author[orcid=0000-0003-1225-3462,gname=Callum,sname=Dewsnap]{Callum Dewsnap}
\affiliation{Department of Physics \& Astronomy, The University of Western Ontario, London, ON, Canada}
\email{cdewsnap@uwo.ca}

\author[orcid=0000-0003-2767-0090,gname=Pauline,sname=Barmby]{Pauline Barmby}
\affiliation{Department of Physics \& Astronomy, The University of Western Ontario, London, ON, Canada}
\affiliation{Institute for Earth \& Space Exploration, The University of Western Ontario, London, ON, Canada}
\email{pbarmby@uwo.ca}

\author[orcid=0000-0001-6217-8101,gname=Sarah C.,sname=Gallagher]{Sarah C. Gallagher}
\affiliation{Department of Physics \& Astronomy, The University of Western Ontario, London, ON, Canada}
\affiliation{Institute for Earth \& Space Exploration, The University of Western Ontario, London, ON, Canada}
\email{sgalla4@uwo.ca}

\correspondingauthor{Pauline Barmby}
\email{pbarmby@uwo.ca}

\begin{abstract}

 The properties of galaxies that host active galactic nuclei (AGNs) offer clues to the conditions that enable accretion onto supermassive black holes. A galaxy’s morphology can provide significant information on the evolutionary processes underway. By modeling  the surface-brightness profiles of galaxies that host AGN, we can characterize physical properties of the host galaxy. With a sample of sources across a wide range of redshifts, we can investigate how both the supermassive black hole and galaxy evolve over time.  Applying the best practices from previous work, we perform S\'ersic + point source fits of 96 X--ray-selected AGN in the Extended Groth Strip  across up to 13 bands of James Webb Space Telescope and Hubble Space Telescope imaging using \textsc{AstroPhot}. We examine both parametric (S\'ersic) and nonparametric methods in tandem to best determine the physical properties of the host galaxy. After separating the light of the host galaxy and the AGN, we perform independent CIGALE model fits to photometric spectral energy distributions to estimate physical properties including the host galaxy stellar mass and star formation rate, and AGN disk luminosity. In the stellar-mass/star-formation-rate plane, the AGN host galaxies align with the population of massive non-AGN galaxies in the control sample across all redshifts. The AGN disk luminosities result in ratios of bolometric to X-ray flux within the expected range, 5--20:1.  We provide an evaluation of the limitations of S\'ersic modeling to obtain galaxy properties and provide recommendations on best practices for multi-band AGN+host galaxy image decomposition with James Webb Space Telescope imaging.

\end{abstract}

\keywords{\uat{AGN host galaxies}{2017} --- \uat{Active galaxies}{17} --- \uat{Active galactic nuclei}{16} --- \uat{Galaxies}{573} --- \uat{Galaxy morphology}{582}}

\section{Introduction} \label{sec:intro2}

Understanding the relationship between an active galactic nucleus (AGN) and its host galaxy is among the key questions in extragalactic astronomy. As AGNs and their hosts co-evolve in time \citep{Kormendy2013}, understanding the ways in which they interact is vital.  Feedback from AGN can significantly impact the surrounding interstellar medium, as winds can eject or heat the surrounding dust and gas, limiting both the star formation and accretion onto the central supermassive black hole \citep[SMBH;][]{Fabian2012, Weinberger2018}. Despite the vastly different size and mass scales of galaxies and SMBHs, \citet{Marconi2003} find significant correlation between the mass of the central SMBH and the luminosity and mass of the host galaxy's bulge, as well as the effective stellar velocity dispersion.  These general correlations are well-supported by many observational studies \citep[e.g.,][]{Ferrarese2000,Kormendy2013}.  While galaxy mergers are certainly implicated in morphological transformation and triggered star formation, their influence on AGN activity is less clear.  For example, \citet{Grogin2005} find minimal connection between mergers and AGN activity, suggesting that internal mechanisms drive the AGN \citep[see also][]{Stone2026}. Other studies have identified links between external forces such as mergers and interactions \citep[e.g.,][]{Sanders1988,Ellison2011,Satyapal2014}.  Obtaining the necessary information to unpack the complex interactions between black hole growth and galaxy evolution requires effective methods for measuring the properties of the AGN and the host galaxy independently.

Measuring the morphology of galaxies tells us about the distribution of stars and dust within. But, in order to accurately determine properties of either the AGN or the host galaxy individually, it is necessary to disentangle the light from the two sources. Such a decomposition is non-trivial, especially at higher redshifts where the host galaxies are fainter and have a smaller apparent size, comparable to the point-spread function (PSF) that defines the AGN size in images.
Decomposing the AGN from its host galaxy is typically performed using parametric methods, most commonly with a two component S\'ersic+PSF fit \citep{Haussler2007, Simmons2008, Kim2008, Gabor2009, vanderWel2012, Ding2020, Ding2022, Dewsnap2023}. The S\'ersic profile is often used to model galaxies due to its ability to model the structure of a wide variety of galaxies at both low and high redshifts \citep{deVaucouleurs1948, Sersic1963, Sersic1968, Freeman1970, deVaucouleurs1976, deVaucouleurs1991}.
However, many galaxies are not well-represented by a S\'ersic profile, particularly if they contain  distinct features such as spiral arms, dust lanes, tidal tails or other irregularities.  As image quality increases, such features become more apparent. The S\'ersic index, effective radius, and integrated brightness obtained by a S\'ersic profile fit are highly correlated, and degeneracy limits the degree to which they can be well-constrained \citep{Ji2020, Ji2022}. The central pixels of a galaxy's light profile are important for constraining the S\'ersic index, particularly for smaller or more distant sources. When performing a S\'ersic+PSF fit, the central pixels can be significantly contaminated by residual effects caused by an imperfect PSF model. Such issues result in highly uncertain measures of brightness and structure in many sources.

Galaxy morphology can also be measured using nonparametric methods that make no assumptions about the shape of the brightness profile, providing a more unbiased indicator of morphology \citep{Bershady2000, Conselice2000, Conselice2003, Lotz2004, Lotz2008}. Nonparametric measures cannot account for the presence of an AGN which significantly influences the results. However, \citet{Dewsnap2025} found that the measurement of AGN brightness from a S\'ersic+PSF fit is robust relative to large changes in the S\'ersic profile modeling the host galaxy. Using a PSF model scaled to the AGN brightness, it is therefore possible to perform AGN subtraction on the source, leaving only the host galaxy remaining. Then, nonparametric galaxy measurements can be made without contamination from the AGN.  Furthermore, clean photometric separation of AGN + host galaxy light across wavelengths allow for independent (rather than joint) modeling of the spectral energy distributions (SEDs).

In this paper, we present parametric and nonparametric methods used in tandem to measure and separate AGN and host galaxy light across a broad range of optical and near-IR wavelengths.  This tandem approach mitigates the downsides of each technique when applied in isolation. We perform S\'ersic+PSF decompositions of 96 AGNs between $0.1<z<4$ on up to 13 filters of Hubble Space Telescope (HST) and James Webb Space Telescope (JWST) imaging from the observed-frame optical to near-IR. We then perform an AGN subtraction and make nonparametric measurements of the host galaxy's light distribution, including aperture photometry and concentration. We compare how the measurements change across filters, as well as how the parametric and nonparametric indicators of structure differ. We investigate the spectral energy distribution (SED) of the host galaxies and AGN using the photometry we obtained.

The structure of the paper is as follows. Section \ref{sec:data2} outlines the sample of AGN and the imaging used. Section \ref{sec:methods} covers the PSF construction, the S\'ersic+PSF decompositions, and the nonparametric measurement process. Section \ref{sec:results2} outlines the results of the measurements and evaluates how the parametric and nonparametric measurements compare. In Section \ref{sec:SEDs}, we apply the software CIGALE to model the SED and determine physical properties of the host galaxies and AGNs.

All magnitudes throughout use the AB system \citep{Oke1983}. We adopt a flat $\Lambda$CDM cosmology with $H_0=70\,\si{\kilo\metre\per\second\per\mega\parsec}$, $\Omega_\mathrm{m}=0.3$, and $\Omega_\mathrm{\Lambda}=0.7$.

\section{Data} \label{sec:data2}

\subsection{Imaging} \label{subsec:imaging2}

Performing the initial S\'ersic+PSF decomposition of a high-redshift galaxy requires high spatial resolution and depth. A deep field with ideal data for this is the Extended Groth Strip (EGS), which has been extensively imaged across a significant fraction of the electromagnetic spectrum. In particular,  HST's Cosmic Assembly Near-IR Deep Extragalactic Legacy Survey  \citep[CANDELS;][]{Grogin2011} and the Cosmic Evolution Early Release Science \citep[CEERS;][]{Finkelstein2023} surveys provide 13 bands of optical to near-IR imaging of the region, including ${\sim}100\,\si{\arcminSpelled\squared}$ of data from the Near Infrared Camera (NIRCam) aboard JWST.
The CEERS survey provides 10 pointings of NIRCam imaging of EGS split across two data releases. The first data release, DR0.5, was observed in June 2022 and contains four of the 10 pointings (specifically, CEERS1--3 and CEERS6). The remaining six pointings (CEERS4, CEERS5, and CEERS7--10) were observed in December 2022 and released in DR0.6. All observations were taken across seven filters: F115W, F150W, F200W, F277W, F356W, F410M, and F444W with $5\sigma$ point source depths between 28.2--29.2 mag. See \citet{Bagley2023} for a description of the data reduction process.

The CEERS team provides pixel-aligned imaging of the overlapping CANDELS data. The HST data includes imaging from the Advanced Camera for Surveys (ACS)/Wide Field Channel (WFC) filters F606W and F814W, as well as the HST WFC3/IR filters F105W, F125W, F140W, and F160W. The $5\sigma$ point source depth of the HST ACS/WFC filters are 28.6 and 28.3 mag for F606W and F814W, respectively. The HST WFC3/IR filters have $5\sigma$ point source depth of 26.7--27.4 mag. See \citet{Koekemoer2011} for the data reduction of the HST CANDELS data. Both the JWST and HST imaging are drizzled to a pixel scale of 0.03\arcsec/pix. The JWST and HST data significantly overlap in the regions of sky covered. The imaging in the HST WFC3/IR F105W filter, however, has little overlapping coverage with the JWST observations.

All the HST/CANDELS and JWST/CEERS data used in this paper can be found in MAST: \dataset[https://doi.org/10.17909/T94S3X]{https://doi.org/10.17909/T94S3X},
\dataset[https://doi.org/10.17909/z7p0-8481]{https://doi.org/10.17909/z7p0-8481}.

\subsection{AGN catalog} \label{subsec:agn2}

The AEGIS-XD survey \citep{Nandra2015} imaged the central region of EGS with Chandra and provide a catalog of 937 X-ray point sources alongside their multiband counterparts from the Rainbow Cosmological Surveys Database \citep{Barro2011a, Barro2011b}. X-ray AGN are selected primarily due to the robustness of their detections---X-ray AGN are much less contaminated by other sources, such as inactive, star-forming galaxies \citep{Hickox2009, Donley2008, Donley2012, Civano2012}.

\begin{figure*}
\includegraphics[width=\textwidth]{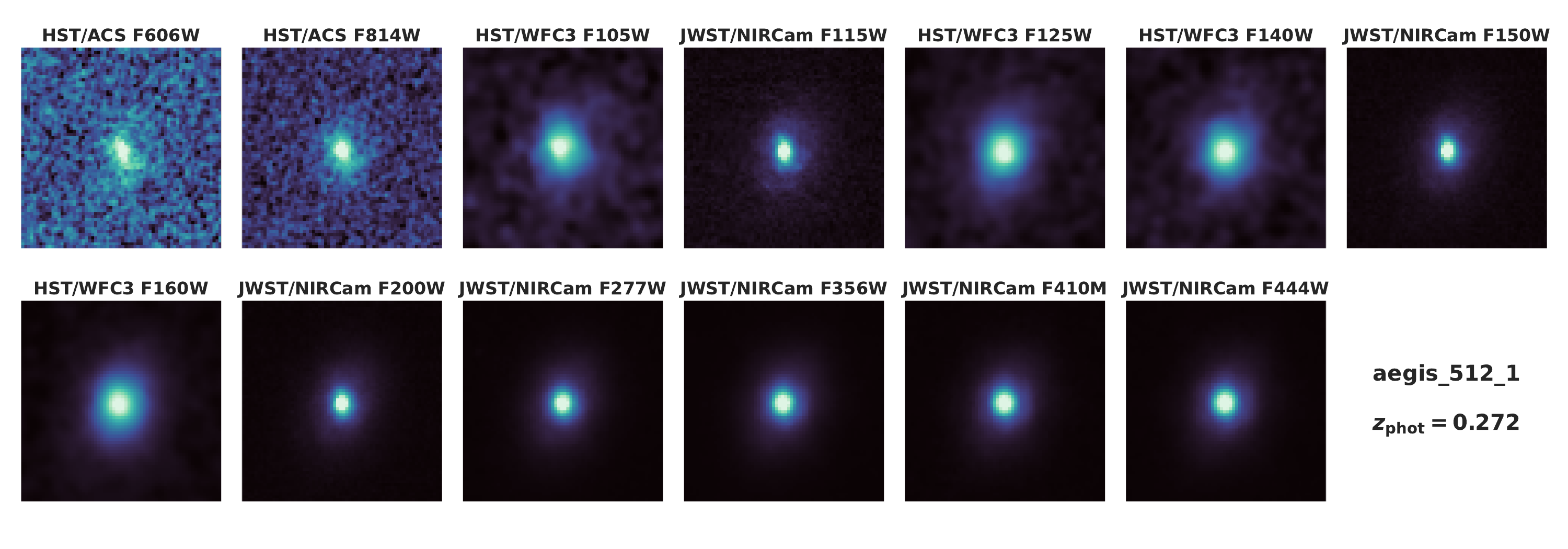}
\caption{Example source imaged across all 13 filters. The example source is \texttt{aegis\_512\_1} with $z_\mathrm{phot}=0.272$. Each cutout is 2{\arcsec} across. These images demonstrate the dominance of the central point source over the galaxy light; changes with wavelength are not readily apparent.}
\label{fig:ex_source}
\end{figure*}

Using the \citet{Nandra2015} catalog, we search for matches to X-ray sources within the CEERS footprint. In our search, we prioritize sources within the bounds of the JWST/NIRCam LW filters (F277W, F356W, F410M, and F444W). Few sources fall within the chip gaps of the JWST/NIRCam SW filters, and so the majority of selected sources are imaged by all seven JWST filters. The vast majority of sources are also imaged by HST (other than HST WFC3/IR F105W), however some sources are too faint to be detected in the shallower HST WFC3/IR bands. Several sources, typically at higher redshift, do not have significant emission in the shorter wavelength filters F606W, F814W, and occasionally F115W. For each source, we visually examine and manually select filters with sufficient S/N such that a S\'ersic+PSF fit is possible. For each accepted filter, we generate a $255\times255$ pixel (7.65\arcsec$\times$7.65\arcsec) cutout centered on the position of the optical counterpart listed in the catalog. 
Figure \ref{fig:ex_source} shows an example of a source imaged across all 13 filters.

\begin{figure}
\centering
\includegraphics[width=\columnwidth]{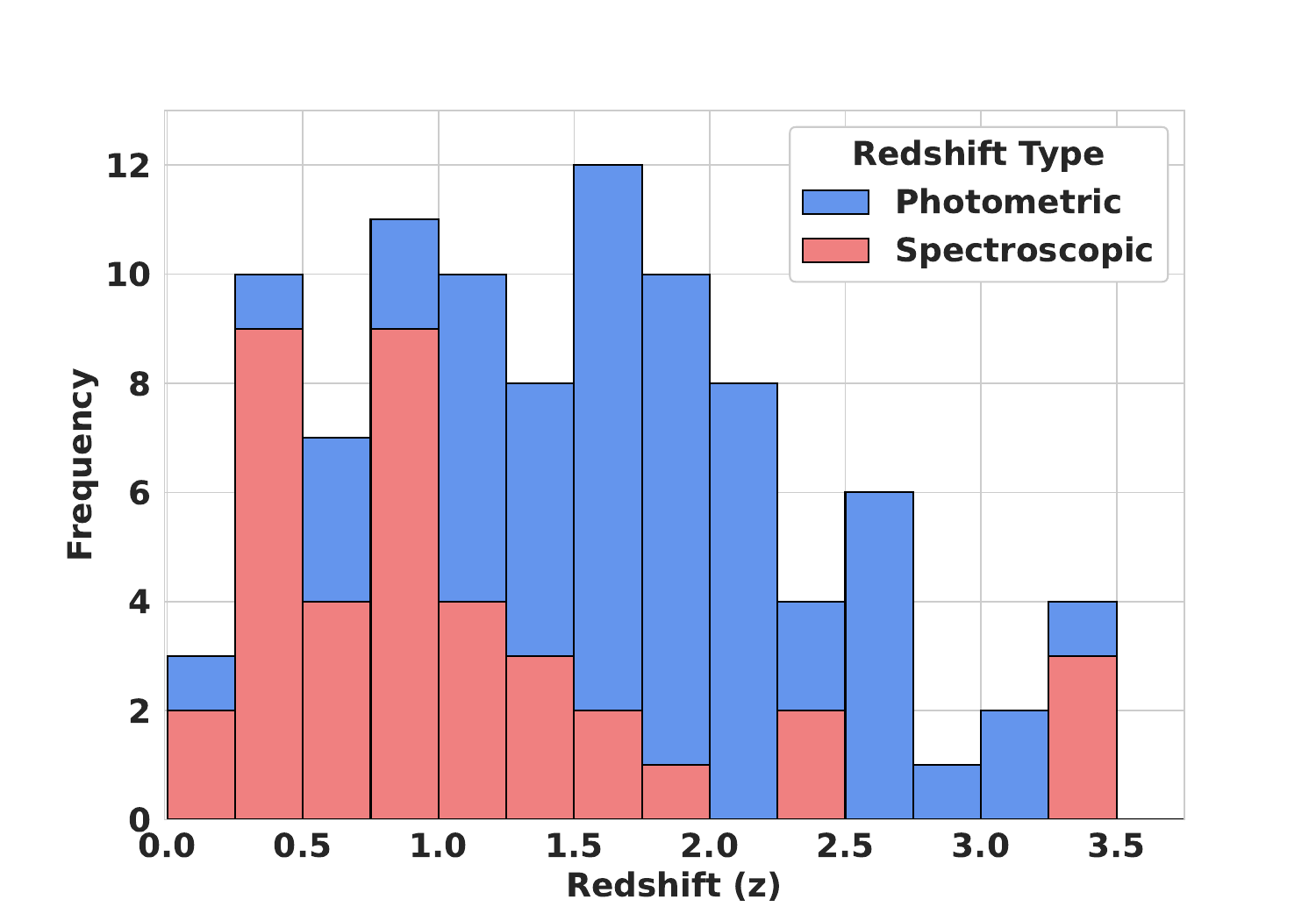}
\caption{Distribution of the redshifts for the 96 selected sources. The blue and red bars indicate  photometric and spectroscopic redshifts, respectively.}
\label{fig:redshifts}
\end{figure}

Our final sample contains 96 sources. Of these, 39 have spectroscopic redshift measurements above the confidence threshold set by \citet{Nandra2015}, and the remaining 57 have photometrically determined redshifts. The spectroscopic redshifts are determined from observations from the Keck and MMT Observatories, whereas the photometric redshifts are from multiband spectral energy distribution fits of up to 35 bands of UV to mid-IR imaging from the Rainbow survey \citep{Barro2011a, Barro2011b}. See \citet{Barro2011b} for a description of the photometric catalog. Figure \ref{fig:redshifts} shows the distribution of redshifts for the 96 sources.
The sample contains five sources for which the position of the AGN is ambiguous due to the presence of two bright, centrally concentrated sources near the coordinates provided by \citet{Nandra2015}. For each of these sources, we perform  analyses twice, once for each possible AGN position.

\section{Methods} \label{sec:methods}

\subsection{PSF Construction} \label{subsec:psfs}

To construct  PSFs, we follow a process similar to those outlined in \citet{Dewsnap2025} and \citet{Zhuang2024}. We create a point source catalog by applying \textsc{SExtractor} \citep{Bertin1996} on each of the CEERS mosaics. In particular, we search for sources that are sufficiently point-like ($\texttt{CLASS\_STAR} \geq 0.9$), do not trigger any significant \textsc{SExtractor} flags ($\texttt{FLAGS} \leq 2$), are not elongated ($\texttt{ELONGATION} < 1.5$), and have a FWHM within 30\% of the PSF value reported by \citet{Finkelstein2023}. We remove any source that coincides with sources in the X-ray catalog outlined in Section \ref{subsec:agn2}. We also require that each candidate point source be detected and accepted in at least three of the filters. This process leaves us with 499 unique point source candidates across all 13 filters. This count ignores duplicate point sources (e.g., a point source which appears in F115W, F150W, and F200W is one source, not three).

After the preliminary \textsc{SExtractor} run, we visually inspect every accepted candidate point source in all available filters. We remove any sources too close to the edge of the imaging, saturated sources, or blended sources not caught by \textsc{SExtractor}. Any sources that are not sufficiently high S/N are removed from the sample as are any sources located in the low S/N out-of-field regions of the NIRCAM Wide Field Slitless Spectroscopy pointings \citep[for details see][]{Finkelstein2025} in the JWST/NIRCam F115W and F356W imaging. Following previous work \citep{Dewsnap2025}, we removed from the sample a number of bright HST/ACS F606W and F814W point sources that showed significant aberration in their core, which significantly affect the resulting PSF. A limited few sources that were not flagged as accepted by \textsc{SExtractor} were manually added to the catalog.
After the manual search, we have a final total of 310 unique point sources---111 from DR0.5 and 199 from DR0.6. For each point source, we create a cutout in each of the accepted filters. Any neighboring sources to the point source within the cutout are masked out.

\begin{table}
\begin{center}
\begin{tabular}{l r c} 
\hline
\multicolumn{1}{c}{Filter} & Number of     & Model Size\\
                           & Point Sources & (\arcsec)\\
\hline
    JWST/NIRCam F115W & 146 & 3.975$\times$3.975 \\
    JWST/NIRCam F150W & 136 & 3.975$\times$3.975 \\
    JWST/NIRCam F200W & 139 & 3.975$\times$3.975 \\
    JWST/NIRCam F277W & 151 & 6.015$\times$6.015 \\
    JWST/NIRCam F356W & 172 & 6.015$\times$6.015 \\
    JWST/NIRCam F410M & 176 & 6.015$\times$6.015 \\
    JWST/NIRCam F444W & 169 & 6.015$\times$6.015 \\
\hline
    HST/ACS F606W & 210 & 4.875$\times$4.875 \\
    HST/ACS F814W & 232 & 4.875$\times$4.875 \\
    HST/WFC3 F105W & 47 & 9.555$\times$9.555 \\
    HST/WFC3 F125W & 279 & 9.555$\times$9.555 \\
    HST/WFC3 F140W & 230 & 9.555$\times$9.555 \\
    HST/WFC3 F160W & 278 & 9.555$\times$9.555 \\
\hline
\end{tabular}
\caption{PSF construction details.\label{tab:point_sources}}
\end{center}
\end{table}

Table \ref{tab:point_sources} provides a breakdown of the number of point sources by filter. Note that the HST/WFC3 F105W filter contains fewer point sources than the other filters; this is due to the significantly smaller area of the F105W imaging relative to the other filters. The HST images generally provide more usable point sources than JWST image because the HST imaging has fewer saturated stars compared to JWST and tends to have more complete coverage of the field.
Our PSF catalog contains more sources than the catalog of \citet{Dewsnap2025} to reflect their finding that including lower S/N candidates did not adversely influence the constructed PSF.

We construct PSFs using \textsc{photutils} \citep{Bradley2022}. For each filter, we create three PSFs using three groups of point sources: a global sample, sources from DR0.5, and sources from DR0.6. The three PSFs are constructed both as a test of the robustness of the model and for further use in the surface brightness fits outlined in Section \ref{subsec:fits}. Note that the CEERS DR0.6 observations were taken with JWST rotated \SI{180}{\degree} relative to its position in DR0.5; thus, we rotate the JWST DR0.6 point sources \SI{180}{\degree} during the PSF construction process. During any fits of DR0.6 data, we rotate the PSF back to match the proper orientation of the observations.

\begin{figure*}
\includegraphics[width=\textwidth]{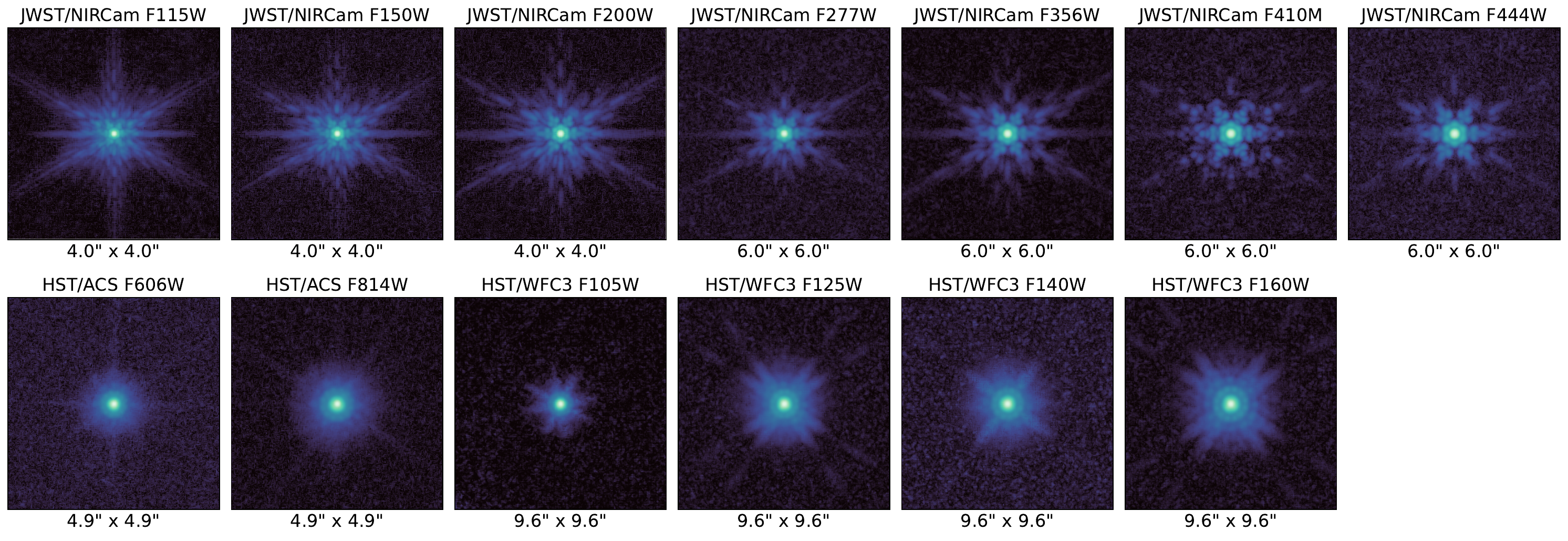}
\caption{Global PSF model for each of the 13 JWST and HST filters used in CANDELS/CEERS imaging of the EGS. Each PSF model has a pixel scale of 0.015\arcsec/pix.}
\label{fig:psfs}
\end{figure*}

We pass each group of point sources into \textsc{photutils} to create the PSF models. We create the models with an oversampling factor of 2. We set $\texttt{maxiter}=5$ and \texttt{norm\_radius} to vary based on image resolution. Due to the limited sample of F105W point sources, we lower the value of \texttt{max\_iter} to 3. See Table \ref{tab:point_sources} for the angular size of the generated models. Figure \ref{fig:psfs} shows the Global PSF for each filter. We measure the FWHM of the PSFs following the process outlined by \citet{Zhuang2024} and \citet{Dewsnap2025}. Our PSF FWHM values are consistently 8--12\% narrower than the values reported by \citet{Finkelstein2023}; \added{while the reason for this is unknown, we note that other studies \citep{Dewsnap2025, Zhuang2024, Wang2024} find results similar to ours and
\citet{Ding2022} find that PSFs generated by the star stacking method outperform the narrower simulated PSFs in studies of quasar host galaxies.
The in-depth analysis by \citet{Zhuang2024} on the effects of PSF mismatch finds that a broader PSF model can mistakenly include the central region of the host galaxy, leading to an overestimation of the AGN flux and underestimation of the host’s Sérsic index. A narrower PSF has the inverse problem--the outer regions of the AGN emission fall outside the PSF model, leading to an overestimated host Sérsic index and underestimation of the AGN flux.}

\subsection{S\'ersic+PSF Decomposition} \label{subsec:fits}

Using the cutouts of the X-ray counterparts discussed in Section \ref{subsec:agn2} and the PSF models from Section \ref{subsec:psfs}, we apply the fitting software \textsc{AstroPhot} \citep{Stone2023} to fit the sources following the process outlined by \citet{Dewsnap2025}. For each source, we fit a two-component model to the cutout in each filter. The components are a S\'ersic profile representing the host galaxy, and a point source component representing the AGN. Models in all bands are fit simultaneously, in what \textsc{AstroPhot} calls a `joint fit'.

For each source, we run \textsc{SExtractor} to obtain  initial conditions. For each cutout, we search using a detection and analysis threshold of $3\sigma$ with a minimum area of 10 pixels above the threshold. Since the majority of the sample we are fitting matches that of \citet{Dewsnap2025}, we can use the final results of those fits to initialize our  fits. We use the \citet{Dewsnap2025} parameters as the initial conditions for the JWST/NIRCam F115W components of the joint fit. Then, we use the parameters from \textsc{SExtractor} as the input for the remaining filters. Namely, we use the \texttt{X\_IMAGE} and \texttt{Y\_IMAGE} for the position of the host galaxy, \texttt{XPEAK\_IMAGE} and \texttt{YPEAK\_IMAGE} as the position of the AGN, $\texttt{THETA\_IMAGE}+90\si{\degree}$ as the position angle, $1/\texttt{ELONGATION}$ for the axis ratio, \texttt{FLUX\_AUTO} with a Kron factor of 2.5 for the total brightness (AGN + host), and \texttt{FLUX\_RADIUS} with \texttt{PHOT\_FLUXFRAC} set to 0.5 for the effective radius. We then use the AGN-to-host ratio and S\'ersic index of the \citet{Dewsnap2025} fits to estimate the remaining initial conditions; namely, the separate AGN and host magnitudes and S\'ersic index. While it is expected that the AGN-to-host ratio and S\'ersic index will vary as a function of wavelength, they act as suitable initial conditions in most cases.

Each fit is performed across all filters simultaneously using \textsc{AstroPhot}'s joint-modeling capabilities. The position of the AGN is held  fixed across all 13 filters, using the JWST NIRCam F277W filter as reference. The center positions of the host and neighboring galaxies are allowed to vary within 5 pixels of the F277W value, as the geometric center can change depending on rest frame wavelength. The F277W filter was chosen as it struck the best balance of S/N, visible extended structure within the host galaxy, and clear central bulge and point source. The galaxy effective radius is also held fixed within 1--127 pixels (0.03--3.81\arcsec), where the upper limit is the radius of the image cutout.
\added{Because the F115W band has the highest resolution and deepest 5$\sigma$ point source depth, including the fainter, lower resolution bands in fitting does not necessarily improve the F115W band results. However, including the F115W fits in the multi-band fitting does improve the fitting of the remaining bands: our experiments showed that in fits of the lower resolution band images individually, the AGN flux is often incorporated into the galaxy with a higher S\'ersic index value, resulting in a failed decomposition. However, the presence of more well-behaved fits in the minimization process can give the necessary ``kick'' to the failed fits, resulting in a successful decomposition.}

The fit results can be significantly influenced by the presence of neighboring sources. We visually inspect every source to determine which neighbors are likely to contaminate the fits. For more distant neighbors, we apply a mask obtained from a \textsc{SExtractor} segmentation map. For more significant neighbors, we add an additional S\'ersic profile to model the neighboring source to minimize the contamination. The neighbors' initial conditions are taken from \textsc{SExtractor} with an initial S\'ersic index of 2. The neighbors that are selected to be fit or masked are not necessarily consistent across all bands---some sources that are disruptive in one band are not in another.

For each source, we perform fits with each of the three PSFs. The choice of PSF is the largest source of \added{empirical} uncertainty for the measurement of the AGN magnitude, which remains robust regardless of strong variations in the associated S\'ersic profile  \citep{Dewsnap2025}. The parameters of the host galaxy, the S\'ersic index and the effective radius, are significantly influenced by both the PSF and the fitting process. As such, we use both the uncertainties reported by \textsc{AstroPhot} and the difference in the fit results between PSFs to estimate the uncertainty of our best-fit measurements \citep{Kimbro2025}.  \added{An additional source of uncertainty is the unknown contribution of any compact  source of stellar light such as a nuclear star cluster to the unresolved PSF emission.  In this case, the contribution of stellar contamination to the PSF light is likely a function of wavelength, both because of the increasing PSF FWHM at longer wavelengths as well as the different shape of the stellar vs. AGN SEDs.  This possibility will be examined further in \S\ref{sec:SEDs}.}

After running each fit, we examined the results to determine if we obtained a reasonable model. To determine if a model is reasonable, we investigate whether the parameters reflect the observable trends between bands (e.g., brightness, degree of disc/bulge dominance, size, AGN-to-host ratio, etc.) and if the results are relatively robust to the choice of PSF. If a fit is determined to be unreasonable, we manually adjust the initial conditions and attempt the fit again.

\subsection{Nonparametric Measurements}

After we complete the S\'ersic+PSF fits, we take the best-fit AGN measurements and perform an AGN subtraction for each source. We subtract any neighbors that were fit and apply the existing masks for any other nearby sources. What remains is an image of only the  host galaxy with as little contamination as possible.
We apply the software \textsc{statmorph} \citep{Rodriguez-Gomez2019} to the AGN-subtracted images to calculate the concentration and elliptical half-light radius of each source across all filters. We do not calculate any other morphometrics, such as smoothness or asymmetry, due to significant residual effects from the AGN subtraction. \textsc{statmorph} reports the Petrosian radius ($R_\mathrm{Petro}$; \citealt{Petrosian1976}), ellipticity, and position angle of the source. Using these, we construct an aperture of size $2 \times R_\mathrm{Petro}$ with which we perform aperture photometry on the AGN-subtracted host galaxy image.

\textsc{statmorph} uses the concentration definition from \citet{Bershady2000}:
\begin{equation}
    C = 5 \log \frac{R_{80}}{R_{20}},
\end{equation}
where $R_{80}$ and $R_{20}$ are the circular apertures containing 80\% and 20\% of the galaxy's total light, respectively. \textsc{statmorph} takes the flux within $1.5 \times R_\mathrm{Petro}$ as the total light for the concentration calculation.
Table~\ref{tab:morph_output} gives the AGN and host galaxy magnitudes from the S\'ersic+PSF fits and the  \textsc{statmorph} measurements for each source in every filter.

 \begin{rotatetable*}
 \begin{deluxetable*}{llrrrrrrrrrrrrrrr}
 \tabletypesize{\scriptsize}
 \tablecaption{Profile fits and nonparametric measurements
 \label{tab:morph_output}}
 \tablehead{
 \colhead{Source} &  \colhead{Filter} &  \colhead{AGN mag} &  \colhead{-unc} &  \colhead{+unc} &  
 \colhead{Host mag} &  \colhead{-unc} &  \colhead{+unc} &  
 \colhead{Host $n$} &  \colhead{-unc} &  \colhead{+unc} &  
 \colhead{Host $R_e$} &  \colhead{-unc} &  \colhead{+unc} &  
 \colhead{Sub. mag} &  \colhead{$C$} &  \colhead{$R_{\rm Petro}$}}
 \startdata
 AEGIS\_616\_1 & F115W & 24.303 & 0.057 & 0.021 & 20.964 & 0.016 & 0.014 & 3.060 & 0.012 & 0.017 & 10.23 &  0.03 &  0.03 & 21.091 & 4.02 & 20.62 \\
 AEGIS\_616\_1 & F150W & 23.713 & 0.057 & 0.018 & 20.590 & 0.013 & 0.013 & 2.865 & 0.011 & 0.012 &  9.88 &  0.02 &  0.02 & 20.695 & 3.95 & 20.57 \\
 AEGIS\_616\_1 & F200W & 23.169 & 0.027 & 0.013 & 20.286 & 0.010 & 0.009 & 2.697 & 0.008 & 0.010 &  9.67 &  0.02 &  0.02 & 20.387 & 3.89 & 20.16 \\
 AEGIS\_616\_1 & F277W & 22.750 & 0.019 & 0.021 & 20.139 & 0.010 & 0.004 & 2.103 & 0.004 & 0.017 &  9.46 &  0.01 &  0.01 & 20.195 & 3.58 & 19.81 \\
 AEGIS\_616\_1 & F356W & 22.337 & 0.037 & 0.009 & 20.452 & 0.013 & 0.006 & 1.884 & 0.005 & 0.021 &  9.88 &  0.01 &  0.01 & 20.492 & 3.44 & 21.10 \\
 \enddata
 \tablecomments{Table~\ref{tab:morph_output} is published in its entirety in the machine-readable format.
       A portion is shown here for guidance regarding its form and content.
       Radius measurements $R_e$ and $R_{\rm Petro}$ are in (0.03 arcsec) pixels. 
       }
\end{deluxetable*}
\end{rotatetable*}

\section{Results \& Discussion} \label{sec:results2}

\subsection{Results of the S\'ersic+PSF Decomposition} \label{sebsec:fits}

\begin{figure*}
\centering
\includegraphics[width=0.8\textwidth]{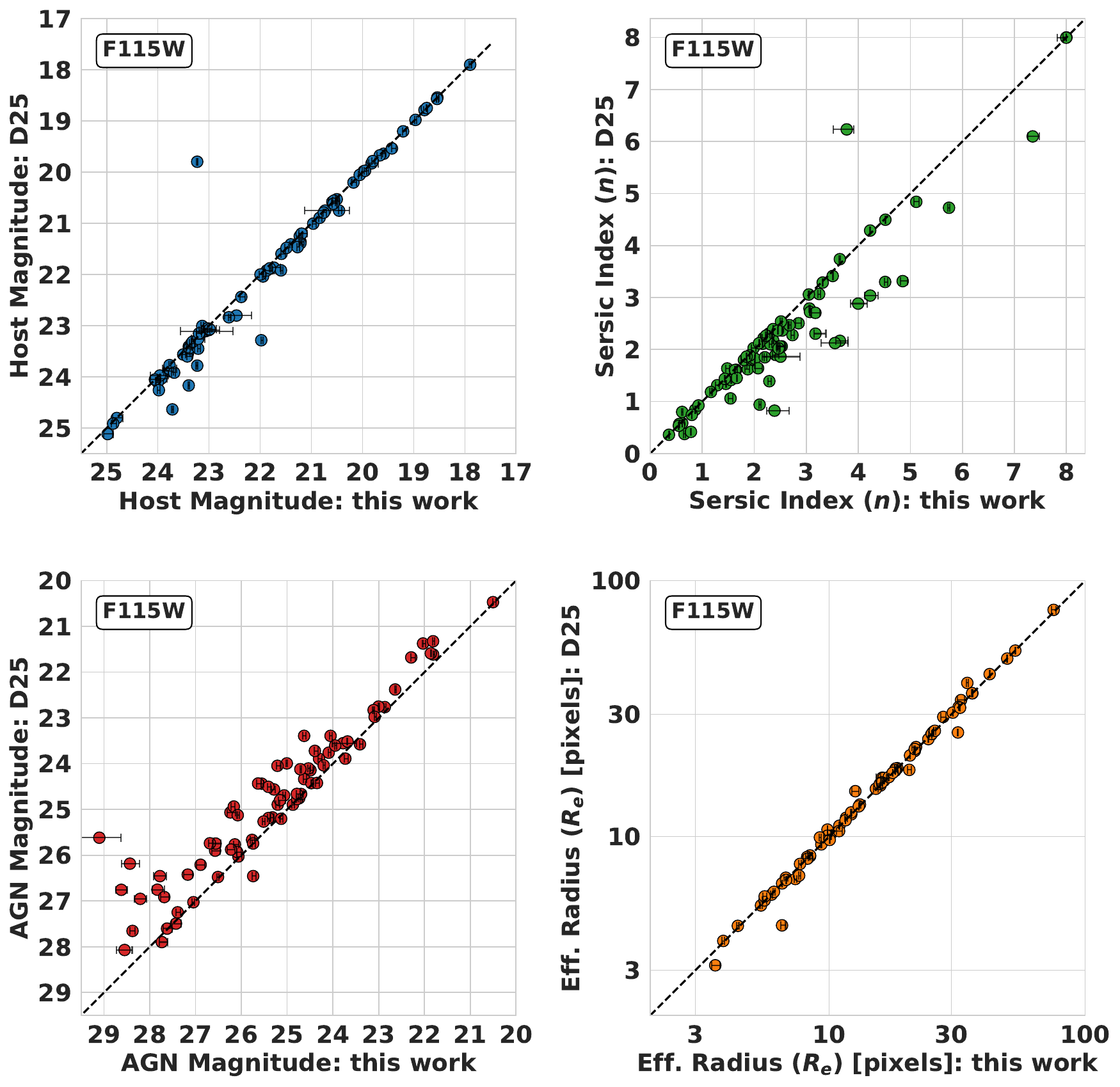}
\caption{Comparison of the JWST F115W fit results from \citet{Dewsnap2025}, denoted D25,  to the results of this work. 
The host magnitude and effective radius measurements generally agree, but this is less the case for the AGN magnitude and S\'ersic index.}
\label{fig:PhD1_vs_PhD2}
\end{figure*}

To test the robustness of the S\'ersic+PSF decomposition, we first directly compare the fits from the JWST F115W filter with the \added{JWST F115W} results from \citet{Dewsnap2025}. 
\added{While the choice of PSF does influence the results of the fits \citep[see Appendix B of][]{Dewsnap2025}, the differences found here are dominated by the change from single- to multi-band fitting.
The fitting methodologies are very similar; the only additions to the fit constraints are those between bands outlined in Section~\ref{subsec:fits} (e.g., holding the AGN position tied together, or limiting variation in host galaxy position). The multi-band fits function identically to the single-band fits in the sense that they minimize a single chi-squared value. However, in the multi-band case, the chi-squared value is a measure across all bands simultaneously.
}
Figure \ref{fig:PhD1_vs_PhD2} shows how the host galaxy magnitude, AGN magnitude, S\'ersic index, and effective radius compare across the two sets of fits. 
While the host galaxy magnitude and effective radius are consistent, the AGN magnitude and S\'ersic index see more significant contrast. 
\added{Fixing the AGN position relative to the F277W image leads to differences in the measured AGN flux, and the Sérsic index is strongly affected by unaccounted-for AGN light.}
In general, the joint modeling of this work resulted in fainter AGN magnitudes, but with higher S\'ersic indices than \citet{Dewsnap2025}---this suggests that the \added{multi-band} fitting process is less successfully disentangling the AGN from the central region of the galaxy, misattributing some AGN light to the central region of the host galaxy. This difference in AGN magnitude is more pronounced for the fainter AGN models. Despite the disagreement, the resulting fits are reasonable overall.

Figure \ref{fig:PhD1_vs_PhD2} also highlights the uncertainties obtained from \textsc{AstroPhot}. Despite using additional fits with other PSF models in an attempt to account for the underlying systematic uncertainties, used alongside the statistical uncertainties reported by \textsc{AstroPhot}, the total uncertainty of the fit parameters is underestimated. The majority of sources report uncertainties of $<1\%$, as opposed to the higher values seen in \citet{Dewsnap2025}.

\begin{figure*}
\centering
\includegraphics[width=\textwidth]{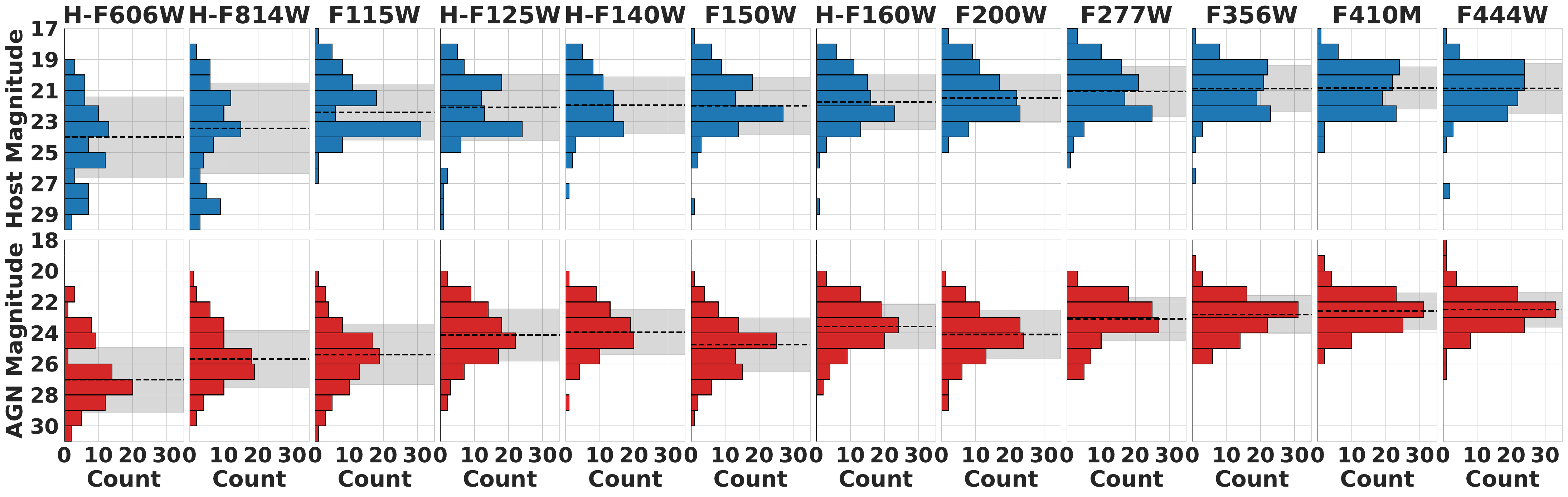}
\caption{Distributions of the host and AGN magnitudes from the S\'ersic+PSF fits across each filter. 
HST filters are indicated with `H-' in the name.
The HST/WFC3 F105W filter is not included due to low sample size. The median value in each panel is indicated with a horizontal dashed line.  Both the host and AGN magnitudes tend to become brighter as the wavelength increases.}
\label{fig:Mag_Hist}
\end{figure*}

Figure \ref{fig:Mag_Hist} shows the distributions of the host galaxy and AGN magnitudes obtained from the S\'ersic+PSF models. We do not include the results from the HST/WFC3 F105W filter here due to the limited sample size. However, the measurements align with the trends described. The general behavior exhibited by each component is similar; as the wavelength of the filter increases, the sources tend to increase in brightness. We also see much tighter distributions at longer wavelengths for both the host and AGN.
In particular, the HST/ACS F606W and F814W measurements show the most widely distributed set of brightness measurements. This is not unexpected---as can be seen in Figure \ref{fig:ex_source}, the two HST/ACS filters experience relatively poor S/N for the extended regions of the sources. For sources at $z>1$, the ACS filters probe the rest-frame NUV and beyond, resulting in clumpy regions of recent star formation that are not well-modeled by a S\'ersic profile. 
\added{Because the host component is not a good representation of the host galaxy in such cases, the decomposition of the AGN from the host is not particularly successful. 
Often the rest-UV images do not contain a clearly defined point source near the fixed position of the AGN component, leading to extremely faint AGN flux measurements.}

\begin{figure*}
\centering
\includegraphics[width=\textwidth]{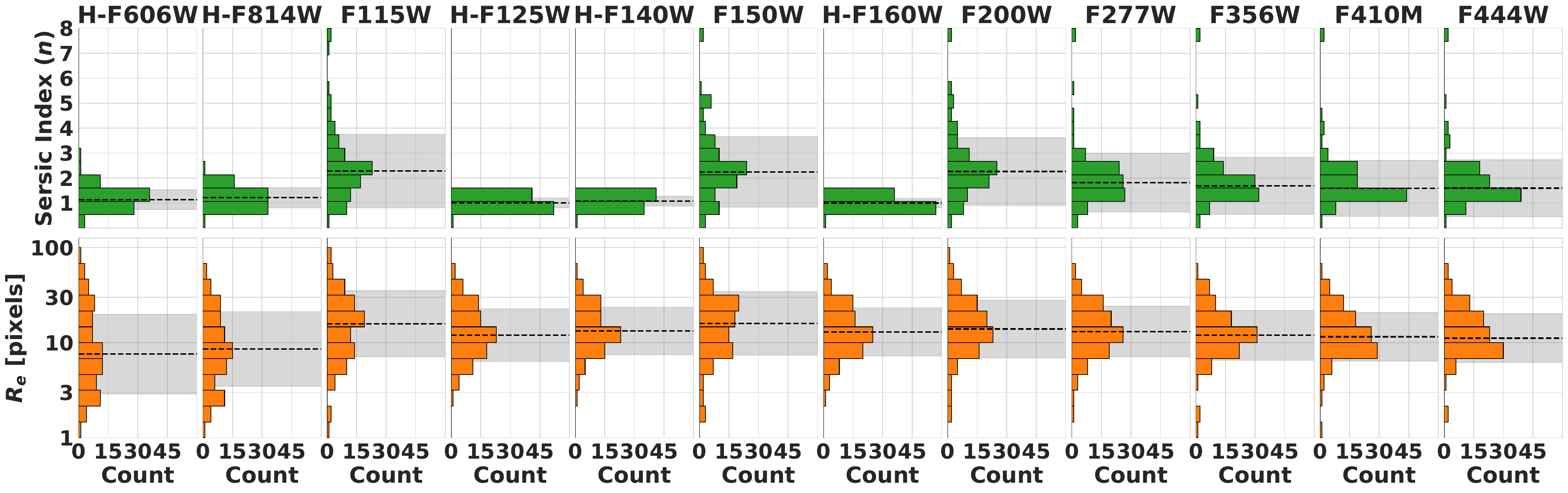}
\caption{Distributions of the S\'ersic index and effective radius from the S\'ersic+PSF fits across each filter. 
HST filters are indicated with `H-' in the name.
The HST/WFC3 F105W filter is not included due to low sample size.  The median value in each panel is indicated with a horizontal dashed line. The HST fits are consistently unable to measure a physically reasonable S\'ersic index.}
\label{fig:Sersic_Re_Hist}
\end{figure*}

Figure \ref{fig:Sersic_Re_Hist} compares the S\'ersic index and effective radius across all filters (again, excluding F105W). We see a striking dichotomy in the S\'ersic index measurements between HST and JWST; while all JWST fits obtain a reasonable distribution within $0.5<n<4$, the HST fits consistently measure S\'ersic indices within $0.5<n<1.5$ for essentially all sources, regardless of wavelength or observed morphological structure. This indicates a consistent inability to determine any information about the light  distribution throughout the host galaxy. Despite this, the radii and brightness measurements generally align with the trends seen using JWST.

The S\'ersic indices  are relatively constant across the JWST filters. This does not fully align with the expectations for the S\'ersic profile's dependence on wavelength. While the S\'ersic index of bulge-dominated $(n_\mathrm{rest, r}>2.5)$ sources are typically relatively constant as a function of wavelength, we expect to see the more disc-dominated sources $(n_\mathrm{rest, r}<2.5)$  have their S\'ersic index increase with wavelength \citep{Vulcani2014}. Since the majority of sources fall within $1<n<2.5$ in the F115W, F150W, and F200W bands, we expect to see the S\'ersic index increase in the F277W, F356W, F410M, and F444W bands.

There are several possible explanations for the discrepancy in the S\'ersic index--wavelength relation. First, it is possible that the systematically lower S\'ersic index found by AstroPhot  \citep{Dewsnap2025} could explain the lower-than-expected values at higher wavelengths. Second, the JWST S\'ersic index distributions are fairly bimodal in structure; the NIRCam SW filters (F115W, F150W, and F200W) have a broader distribution with a peak at higher values and a broader span, whereas the NIRCam LW filters (F277W, F356W, F410M, and F444W) quickly fall to a lower and narrower distribution. The decrease in spatial resolution between the SW and LW channels may result in less accurate fits, similar to how the HST/WFC3 fits consistently failed to a value $n\sim1$. Alongside the jump in resolution, as the wavelength of the images increases, galaxies appear more bulge-dominated. It is likely that the lower resolution and higher central concentration cause the AGN component to be over estimated, artificially lowering the central concentration of the host galaxy's light profile.

Generally, the effective radius of galaxies decreases as a function of observed wavelength. Bulge-dominated sources tend to have their radii decrease much more quickly than disc-dominated sources \citep{Vulcani2014}. Effective radii distributions are fairly constant across filters, with a slight tendency to decrease as the wavelength increases. Since our sources are mostly disc-dominated, this relatively small shift in effective radius matches  expectations. If the S\'ersic indices are systematically lower than expected, as described above, then the degeneracy of $n$ and $R_e$ within the S\'ersic profile would attempt to balance around the host magnitude, resulting in consistently larger effective radii at longer wavelengths \citep{Dewsnap2025}.

\begin{figure*}
\centering
\includegraphics[width=\textwidth]{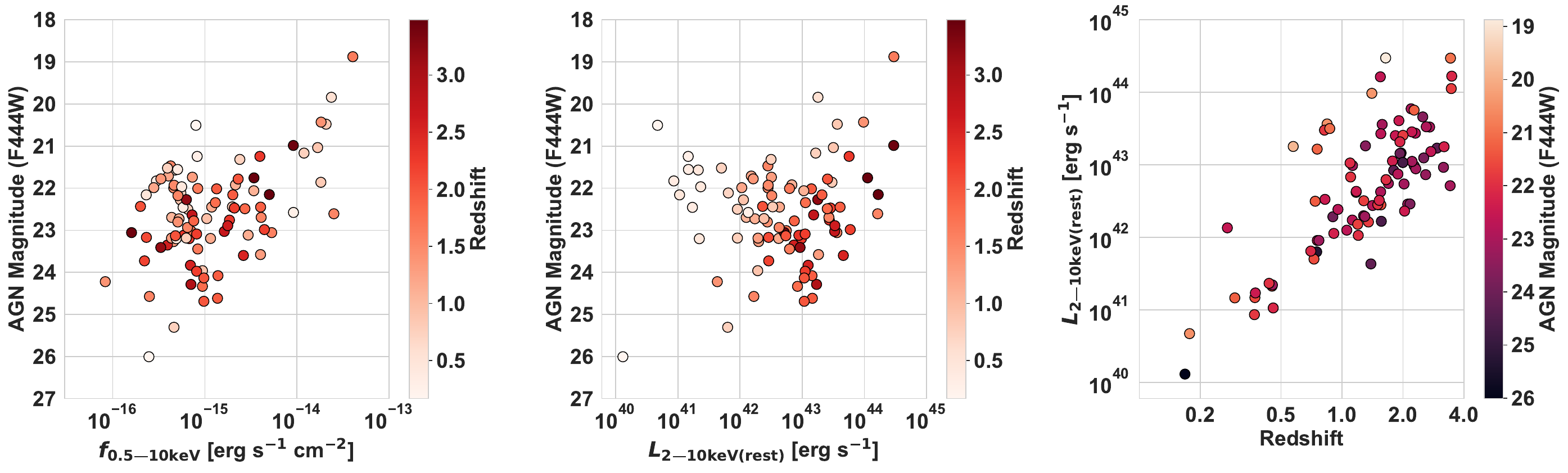}
\caption{Comparison of the AGN apparent magnitude measured in the S\'ersic+PSF fits for the JWST F444W band versus the \SIrange{0.5}{10}{\kilo\electronvolt} flux (leftmost panel) and the rest-frame \SIrange{2}{10}{\kilo\electronvolt} luminosity (middle panel). As expected, the AGN apparent magnitude correlates with the X-ray flux (both originating from the same source), and apparent magnitude is uncorrelated with X-ray luminosity. For reference, the rightmost panel shows the X-ray luminosity as a function of redshift. The JWST F444W band was selected as it is representative of all the JWST filters, while exhibiting the lowest scatter in the AGN magnitude. }
\label{fig:Xray_AGN_Mag}
\end{figure*}

Figure \ref{fig:Xray_AGN_Mag} highlights the relationship between the X-ray flux and luminosity with the AGN magnitude. Following \citet{Nandra2015}, the \SIrange{2}{10}{\kilo\electronvolt} luminosity was calculated using the \SIrange{0.5}{10}{\kilo\electronvolt} flux by applying a K-correction with an assumed power-law spectrum with $\Gamma=1.4$, assuming the flux is unabsorbed.  We only show the results of the F444W fits, but the trends are representative of all JWST filters with the SW filters having more scatter due to the broader magnitude distributions. The AGN (apparent) magnitude measurement correlates with the X-ray flux, particularly at the extremes (the brightest and faintest sources), as would be expected for AGN. There is no correlation of X-ray luminosity with AGN apparent magnitude, as
expected. 

\begin{figure*}
\centering
\includegraphics[width=\textwidth]{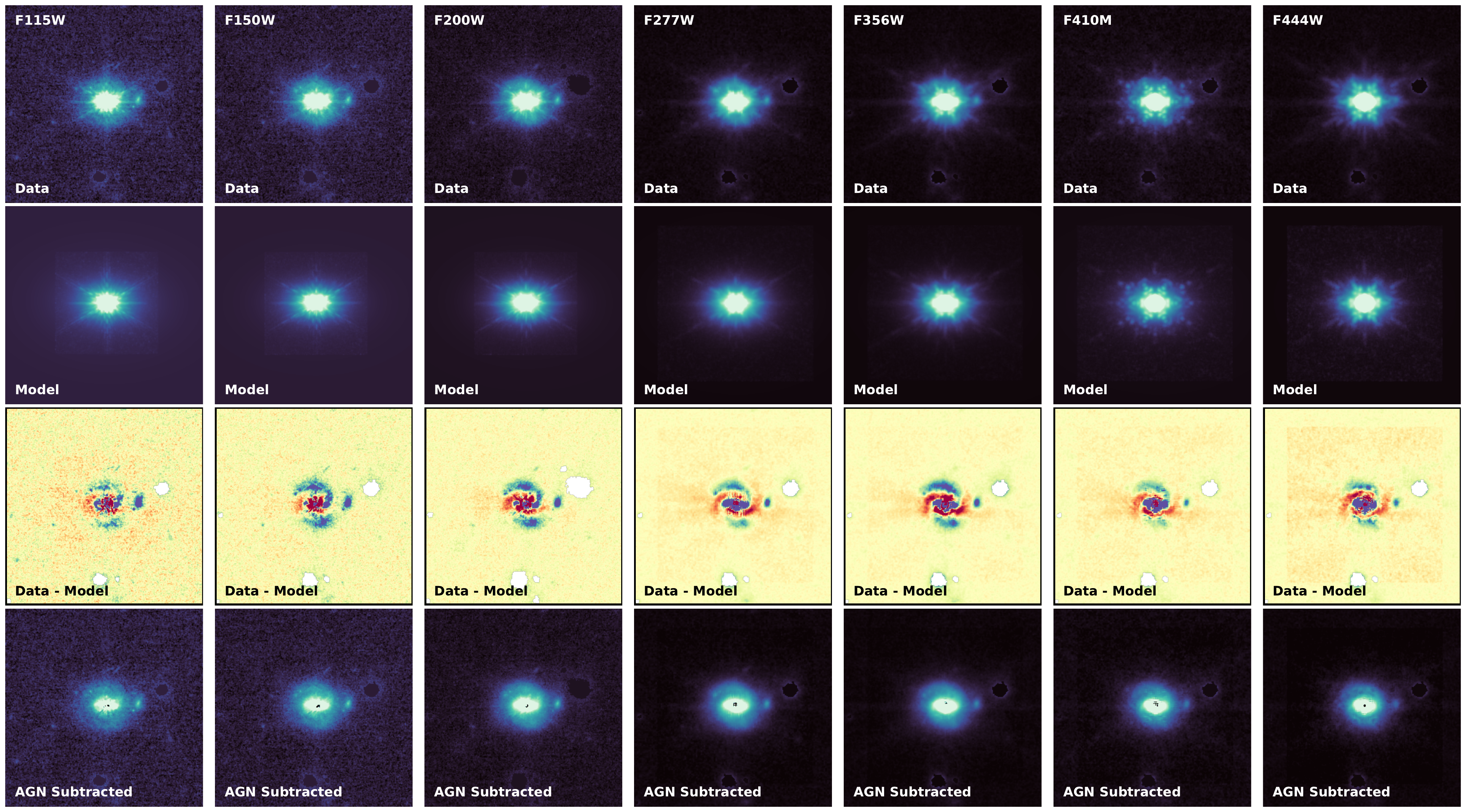}
\caption{Example fit of the source SDSS1420+5300A (Rb\_ID: \texttt{aegis\_742\_1}, $z_\mathrm{spec}=1.644$) across all available JWST filters. Each cutout is 255$\times$255 pixels (\SI{7.65}{\arcsec}$\times$\SI{7.65}{\arcsec}). The top row shows the original image; row 2 shows the best-fitting PSF+S\'ersic model fits. The residuals are shown in row 3.  The galaxy bar and spiral arms are clearly visible after AGN subtraction (row 4); their appearance does not change strongly with wavelength. The companion source to the right of the galaxy (not included in the model fits) decreases in relative brightness with increasing wavelength.}
\label{fig:ex_fit}
\end{figure*}

As an example of the AGN--host decomposition, Figure \ref{fig:ex_fit} shows the results of the JWST fits for the source SDSS1420+5300A (Rainbow ID: \texttt{aegis\_742\_1}, RA: 215.02339290, DEC: +53.01020700). This source is a massive barred spiral ($M_*=10^{11.3}M_\sun$) located at $z_\mathrm{spec}=1.644$. Notably, SDSS1420+5300A hosts a luminous (rest-frame $L_{2-10 \ \mathrm{keV}} = 2.98 \times 10^{44}$~\added{erg~s$^{-1}$}) broad-line quasar with $M_\mathrm{BH} = 10^{9.00\pm0.16} M_\sun$ \citep{Grier2019, Nandra2015, Ding2022, Zhuang2024}. The fits of SDSS1420+5300A demonstrate a successful decomposition of the AGN from the host galaxy. The bar and massive spiral arms are not visible in the original image, but appear very distinctly after subtracting the AGN. The successful subtraction of the PSF viewed in 2D highlights the success of the PSF models across each filter, with no systematic errors across the sources. However, there are clear residual effects near the center of the AGN subtracted image. These anomalous pixels can significantly influence certain morphometric measurements.

\subsection{Galaxy Morphometrics}

\begin{figure*}
\centering
\includegraphics[width=\textwidth]{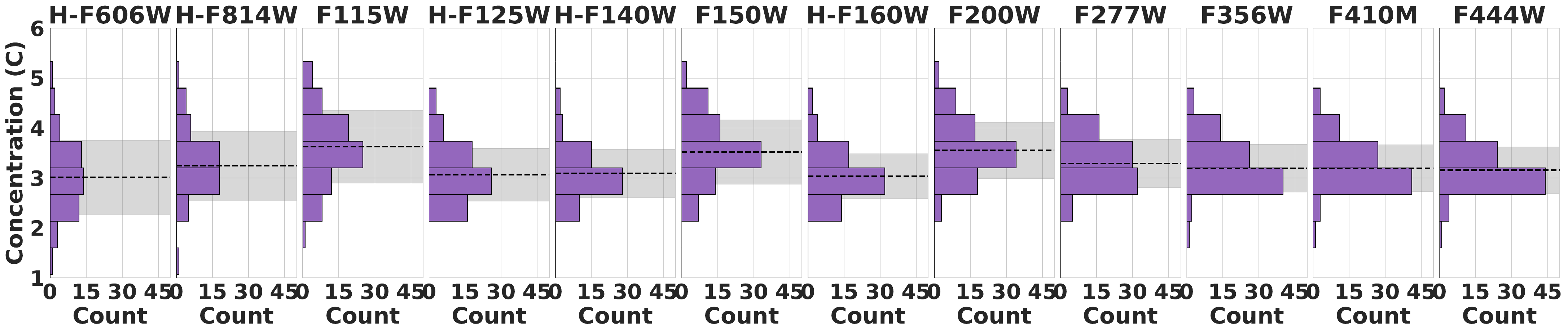}
\caption{Distribution of the concentration measurement of the AGN subtracted images across all filters, excluding HST/WFC3 F105W.  The median value in each panel is shown as a horizontal dashed line.
HST filters are indicated with `H-' in the name.}
\label{fig:concentration_hist}
\end{figure*}

\begin{figure*}
\centering
\includegraphics[width=\textwidth]{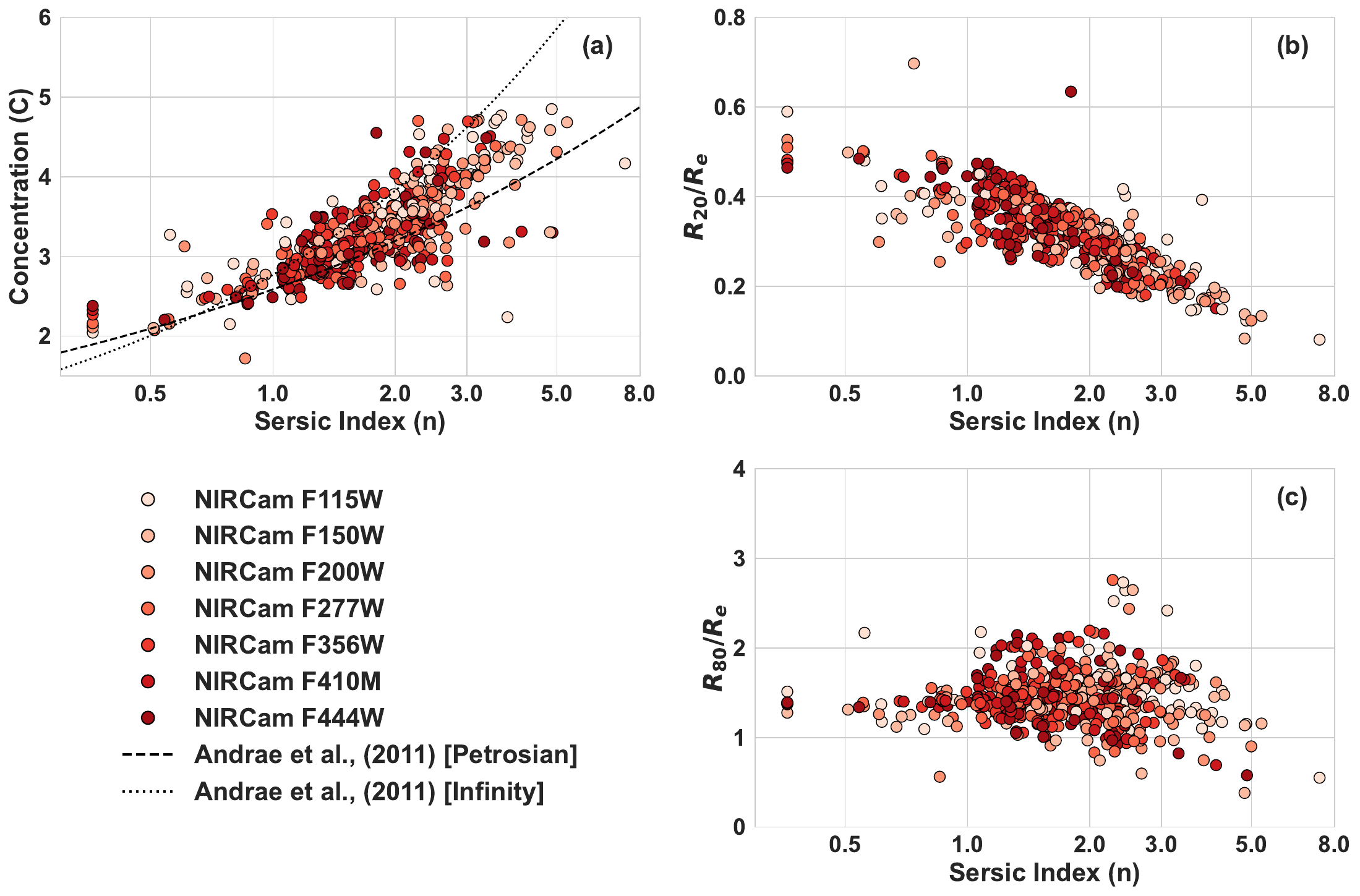}
\caption{The concentration, $C$ (a), $R_{20}/R_e$ (b), and $R_{80}/R_e$ (c) as a function of S\'ersic index for the JWST filters. The black curves represent the relations between the concentration and S\'ersic index given by \citet{Andrae2011} if the S\'ersic profile is integrated to one Petrosian radius ($C_{Petro}$; dashed) or  an infinite radius ($C_{\infty}$; dotted).}
\label{fig:concentration_vs_sersic}
\end{figure*}

Figure \ref{fig:concentration_hist} shows the distribution of the concentration of the AGN-subtracted images. The distribution appears almost as a combination of the S\'ersic index and effective radii distributions---the shorter wavelengths see a wide range of values which tighten as the wavelength increases, however there is a clear disconnect between the HST concentrations compared to JWST, although not as drastic as the S\'ersic index. Similar to the S\'ersic index, we expect that the concentration should increase as the wavelength increases. However, the concentration remains roughly constant, and even decreases between the NIRCam SW and LW filters. This is another indicator that the AGN subtraction from the bulge at longer wavelengths may not be robust.

Since the S\'ersic index and concentration are both descriptors of the shape of the light profile, we expect the values to be correlated. \citet{Andrae2011} provide two approximations of the relationship determined by integrating the S\'ersic profile outwards to both one Petrosian radius and to infinity, resulting in the expressions
\begin{equation}
C_{Petro} \approx 2.586 n ^{0.305}
\end{equation}
and
\begin{equation}
C_{\infty} \approx 2.770 n ^ {0.466},
\end{equation}
respectively.

Figure \ref{fig:concentration_vs_sersic} shows the relationship between the concentration index ($C$), the $R_{20}/R_e$, and $R_{80}/R_e$ values as a function of the S\'ersic index alongside the approximations from \citet{Andrae2011}. We find that the concentration and S\'ersic indices are highly correlated and do mostly align with the approximations. The agreement is consistent across the whole redshift and wavelength range of the sample.

We note that the values determined by \textsc{statmorph} are calculated by measuring the flux out to $1.5R_p$ rather than the $R_p$ of \citet{Andrae2011}. Since we include additional extended light, this results in  increased values of $R_{80}$ and $C$. The systematic offset in the \textsc{AstroPhot} S\'ersic index measurements compared to \textsc{Galfit} reported by \citet{Dewsnap2025} could also explain the offset from the \citet{Andrae2011} expression at $n>1.5$. \citet{Wang2024} find that the concentration can be substantially overestimated ($>{30\%}$ for $n=1$) for highly elliptical ($q<0.5$) sources due to \textsc{statmorph}'s use of circular apertures for $R_{20}$ and $R_{80}$, leading to further deviation from the \citet{Andrae2011} approximation.

\begin{figure*}
\centering
\includegraphics[width=0.8\textwidth]{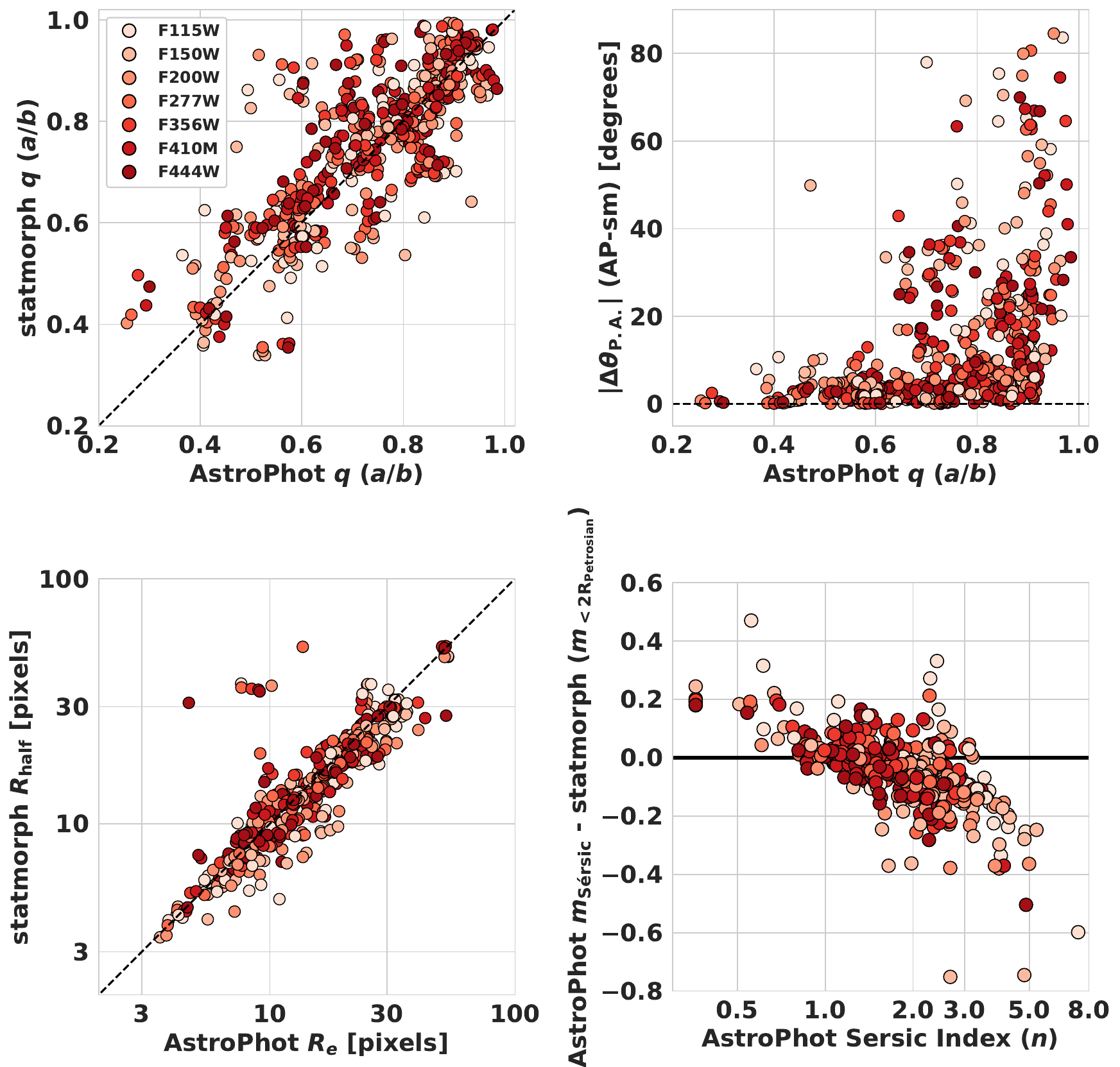}
\caption{Comparison of the axis ratio (upper left panel), the position angle \added{difference} (upper right panel), and the half-light (effective) radius (lower left panel) as reported by \textsc{AstroPhot} and \textsc{statmorph}. The lower right panel shows the difference between the magnitude within two Petrosian radii and the magnitude obtained by integrating the S\'ersic profile to infinity.}
\label{fig:AstroPhot_vs_statmorph}
\end{figure*}

Figure \ref{fig:AstroPhot_vs_statmorph} compares some of the structural information from the S\'ersic+PSF fit to the values extracted from \textsc{statmorph}. Generally, the shape of the profile is consistent. The half-light radius between the two methods agrees outside of a few outliers. For the axis ratio, the strongest disagreement comes from the smaller sources, where the extended structure is much less well-determined. Similarly, the orientation of the source tends to agree outside of smaller or more circular sources, as expected.

Since the flux reported by \textsc{statmorph} is truncated by the aperture, it is expected that the flux from integrating the S\'ersic profile to infinity should be strictly brighter. However, we see a significant population of sources where the Petrosian magnitude is brighter than the S\'ersic magnitude. The majority of these sources are at lower values of the S\'ersic index ($n<2$). Disc-dominated sources such as these tend to host more  features that are not well-modeled by a S\'ersic profile, such as spiral arms, bars, dust lanes, etc. These features can influence the shape of the S\'ersic profile more significantly than the Petrosian aperture. For example, a S\'ersic profile can easily align itself with a bar, prioritizing the central region over the extended structure of the source. Because the aperture photometry measurements make no assumptions about the radial profile of the galaxy, they are not as significantly influenced by deviations from some predefined function. Bulge-dominated sources ($n>2$) are more significantly impacted by the aperture cutout, as higher values of $n$ have more extended wings compared to discs, where the radial profile quickly drops at higher radii. The $2R_{Petro}$ cutout thus excludes significantly more light for higher $n$.

Irregular galaxies tend to fall into the low $n$ population. Because, by definition, they have no clearly defined structure, they are not well represented by a S\'ersic profile. Typically,  a S\'ersic fit will either fit a small fraction of the source (e.g., a single clump) or will result in a larger, flatter profile over a larger region. Either way, the S\'ersic fit is not representative of the source's true brightness. Aperture photometry tends to perform better for such sources, again leading to a number of sources with a brighter Petrosian brightness than S\'ersic.

\section{Spectral Energy Distributions} \label{sec:SEDs}

\subsection{Applying CIGALE}

To extract physical information from the decomposed AGN and host galaxies, we apply the software CIGALE \citep{Boquien2019, Yang2022} to fit SED models to host galaxy and AGN photometry. We apply settings similar to the works of \citet{Magnelli2023} and \citet{Zhuang2024}. For each source, we hold the redshift fixed to the value reported by \citet{Nandra2015}. To model the host galaxy, we use a delayed star formation history with an optional exponential burst, the \citet{Bruzual2003} stellar model, a nebular emission-line model, the modified%
\footnote{\added{The modification applied in CIGALE is the extension of the
attenuation curve ``with the \citet{Leitherer2002} curve between the Lyman break and 150 nm'' \citep{Boquien2019}. CIGALE also allows for this attentuation curve's slope to be varied and a UV bump added; neither option was used in this work.}}  
attenuation law from \citet{Calzetti2000}, and the dust emission model from \citet{Draine2014}. 
For the AGN, we apply the AGN models from \citet{Stalevski2016}. See Table \ref{tab:CIGALE_params} for a description of the parameter grids within each component. Any parameters not listed remained at the default values.

\begin{table*}
\begin{tabular}{l l l}
\hline
Component & Parameter & Value(s)\\
\hline
    \texttt{sfhdelayed} & $\tau_\mathrm{main}$ & 500, 1000, 2000, 5000 Myr \\
                        & $t_\mathrm{main}$    & 500, 1000, 2000, 3000, 4000, \\
                        &                      & 5000, 6000, 7000 Myr \\
                        & $f_\mathrm{burst}$   & 0, 0.1, 0.2, 0.3 \\
     \texttt{bc03} & Initial Mass Function & \citet{Chabrier2003} \\
    \texttt{dustatt\_modified\_starburst} & $E(B-V)$ & 0, 0.005, 0.05, 0.1, 0.2, 0.3,\\
                                          &          & 0.4, 0.5, 0.6, 0.7 \\
    \texttt{dl2014} & $q_\mathrm{PAH}$ & 0.47, 1.12, 1.77, 2.5 \\
                    & $U_\mathrm{min}$ & 5, 12, 20 \\
                    & $\gamma$         & 0.01, 0.05, 0.1, 0.5 \\
    \texttt{skirtor2016} & $\tau_{9.7\si{\micro\meter}}$ & 3, 5, 7, 9, 11 \\
                         & Power-law exponent & 0, 0.5, 1, 1.5\\
                         & Dust density gradient index & 0, 0.5, 1, 1.5\\
                         & Angle between equ.\ plane \& torus edge & 20\si{\degree}, 40\si{\degree}, 60\si{\degree}, 80\si{\degree}\\
                         & $R_\mathrm{outer}/R_\mathrm{inner}$ & 10, 20, 30\\
                         & $i$ & 0\si{\degree}, 30\si{\degree}, 60\si{\degree}, 90\si{\degree}\\
\hline
\end{tabular}
\caption{Settings used for CIGALE fits. The unlisted parameters remained at their default values.\label{tab:CIGALE_params}}
\end{table*}

We applied CIGALE to sources that were successfully fit with both \textsc{AstroPhot} and \textsc{statmorph} and had magnitude measurements for both the host and AGN brighter than $m=30$ in at least four JWST filters. This results in a set of 66 sources to be passed to CIGALE. We apply CIGALE twice to each host galaxy component: once with the S\'ersic photometry measurements and again with the aperture photometry measurements from \textsc{statmorph}. 
Because \added{the mophological} fits significantly underestimate the photometric uncertainty and \textsc{statmorph} does not report uncertainties, we assume a 15\% uncertainty on all photometry measurements \added{based on advice from experienced practitioners (also see below).}%
\footnote{\added{CIGALE's default settings include adding a 10\% relative uncertainty in quadrature to the input photometric uncertainties;
\citet{Jespersen2026} report that CIGALE fits often fail without this `softening'.}}

\begin{figure*}
    \centering
    \includegraphics[width=\linewidth]{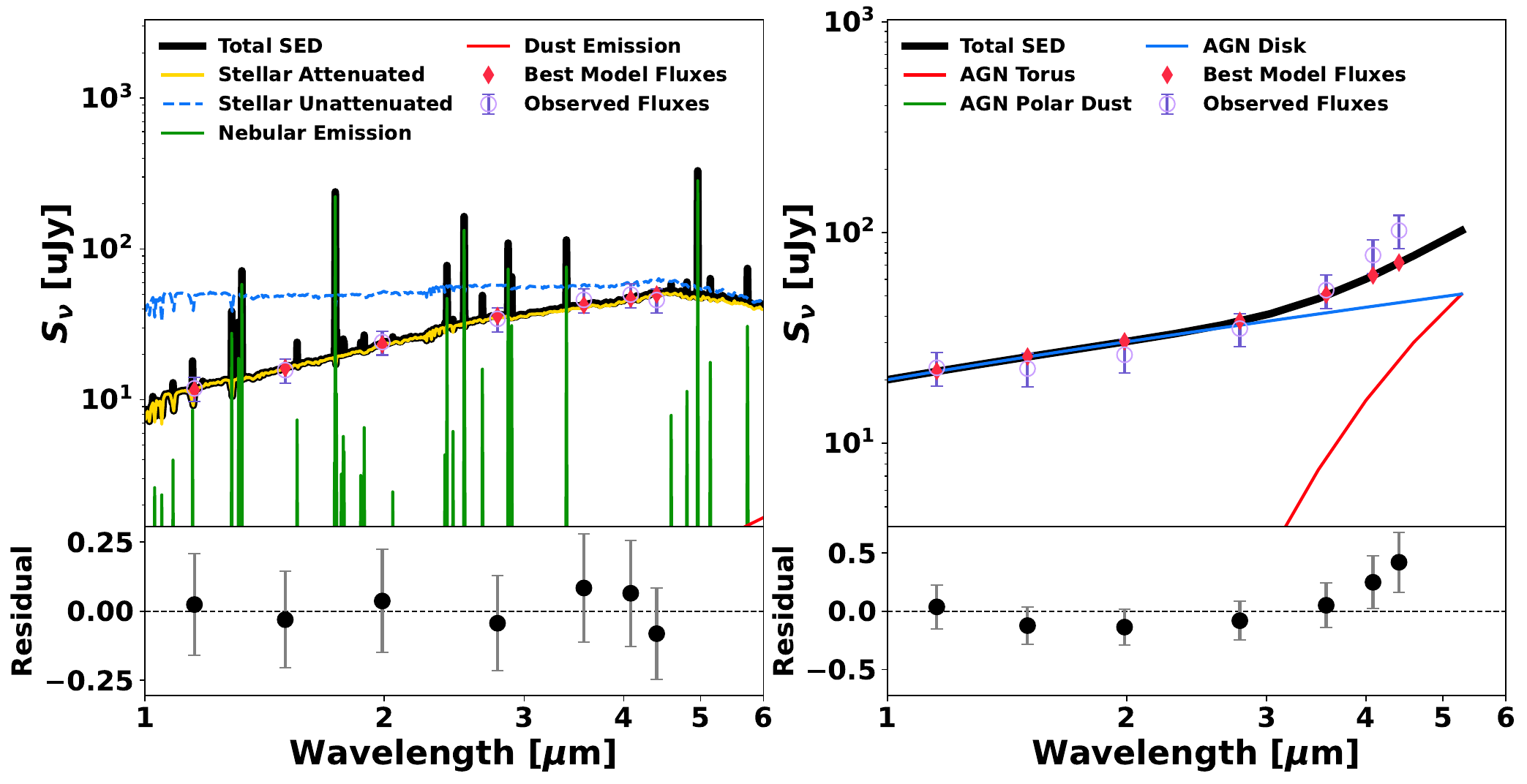}
    \caption{SED fit of both the AGN-subtracted host galaxy (left panel) and the AGN (right) for SDSS1420+5300A (Rb\_ID: \texttt{aegis\_742\_1}, $z_\mathrm{spec}=1.644$). The bottom of each panel shows the relative residual of the fit. This best fit model used S\'ersic models of the host galaxy to calculate the photometric values. See Figure \ref{fig:ex_fit} for the S\'ersic+PSF fits of the source.}
    \label{fig:Ex_SED_Fit}
\end{figure*}

We are generally able to obtain a reasonable fit to host galaxy and AGN SEDs, with median reduced $\chi^2 \lesssim 1$.  
For host galaxy properties, we take the fit with the lower reduced $\chi^2$ to be the best fit. This criterion results in 46/66 sources using the aperture photometry-derived flux measurement as opposed to 20/66 using the S\'ersic fit results.  
When the CIGALE fits using the aperture photometry perform better than the S\'ersic photometry, they tend to significantly outperform. Conversely, the S\'ersic best fits are more similar to the aperture photometry models.
Figure \ref{fig:Ex_SED_Fit} demonstrates an example of the SED model fits, with the contributions of different components highlighted in different colors.
Tables~\ref{tab:SED_fits_glx} and \ref{tab:SED_fits_agn} give the SED model outputs for each of the 66 sources fit by CIGALE.
\added{To address the issue of possible contamination of the AGN SEDs by a compact stellar source such as a nuclear star cluster, we
inspected the AGN model fits and residuals for evidence of a stellar continuum component and found none. 
We experimented with using CIGALE to fit a combination of stellar and AGN SEDs to the AGN photometry and were unable to obtain successful fits and so
conclude that there is no evidence of strong contamination of the AGN light by central stellar populations.}

We find a relation between the size and AGN fraction with the reduced $\chi^2$ of the CIGALE fits. Larger host galaxies with fainter AGN tend to have a lower $\chi^2$ for the host galaxy fit across both the \textsc{AstroPhot} and \textsc{statmorph} CIGALE fits. Similarly, point-like sources with host galaxies that are faint relative to the AGN tend to have lower $\chi^2$ in the AGN CIGALE fits. These results make intuitive sense; minimizing one of the two components leads to a better fit in the other. Sources with both a bright AGN and a large host galaxy or a faint AGN with a small host galaxy exhibit higher uncertainty in their fits.

\input{TabSED}

\subsection{Extracting Physical Information}

\begin{figure}
\centering
\includegraphics[width=\columnwidth]{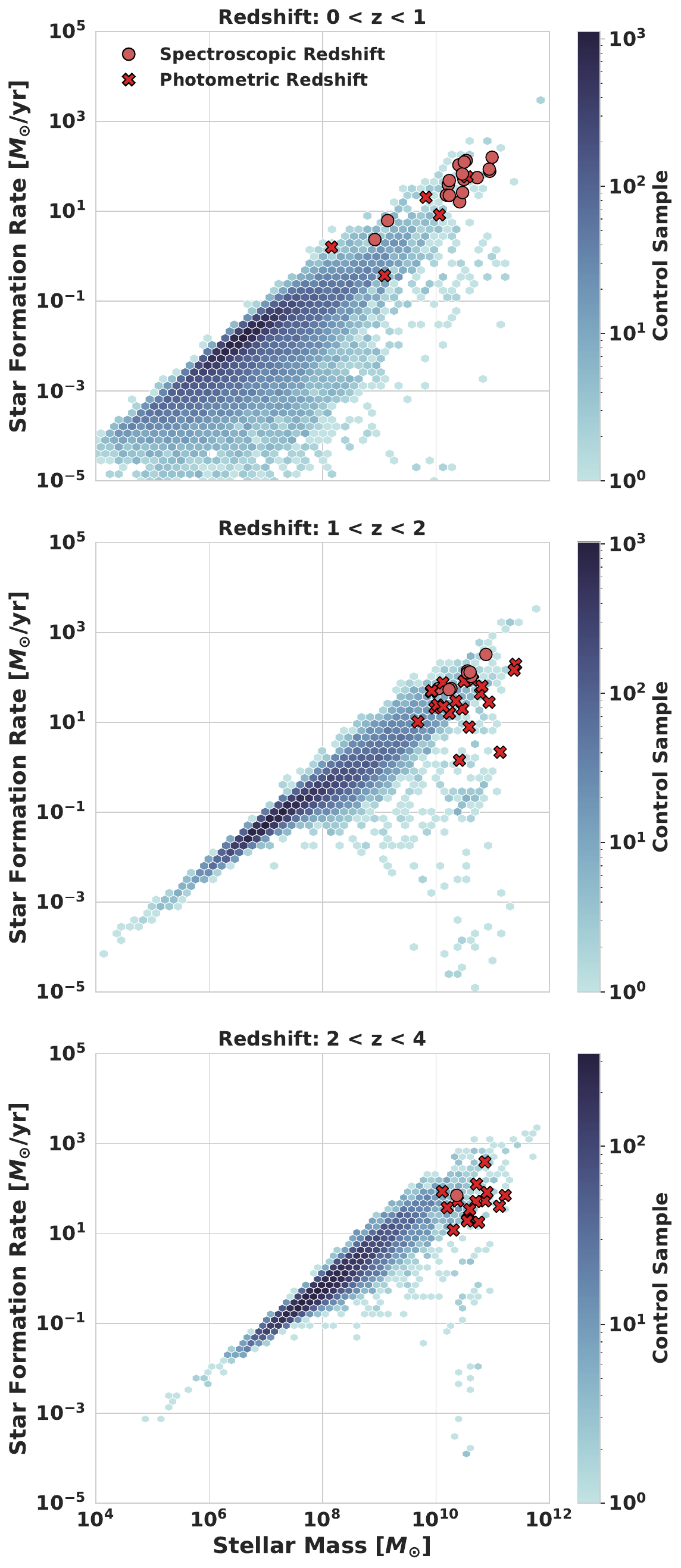}
\caption{Host galaxy SFR compared to $M_*$. Sources are grouped into redshift bins. The red markers indicate the SED fit results, with circles and crosses representing sources which have spectroscopic and photometric redshift measurements, respectively. The blue heat map shows the SFR and $M_*$ for non-AGN galaxies from the CEERS photometric catalog \citep{Cox2025}.  Based on this small sample, the AGN host galaxies have properties consistent with the high-mass end of the galaxy distributions in all redshift bins.}
\label{fig:SFR_vs_Stellar_Mass}
\end{figure}

Figure \ref{fig:SFR_vs_Stellar_Mass} shows the extracted host galaxy star formation rate (SFR) as a function of stellar mass ($M_*$). We also include a comparison with a sample of galaxies without AGN within CEERS. The CEERS photometric catalog \citep{Cox2025} performed SED fits using CIGALE and report the SFR and $M_*$ for ${>}80\ 000$ galaxies between $0<z<14$. For sources within the typical redshift range ($0<z<4$), we find that  AGN host galaxies align with the population of massive non-AGN galaxies in the control sample across all redshifts. At higher redshifts, we find that the host galaxies in our sample may tend to lie in the region of lower SFR massive galaxies, though the numbers are small (15). Furthermore, the majority of higher redshift sources have photometric redshift measurements, leading to more uncertain distance measurements and therefore physical parameters extracted from the photometry. Additionally, the uncertainties of the physical quantities are much more significant ($>$35\%) than those of the input flux (15\%), suggesting that the uncertainties are dominated by systemic effects (such as degeneracies within the model components), rather than our input.
\added{
This is supported by experiments in which we varied (to 0\% and 20\%) the amount of additional photometric uncertainty assumed by CIGALE and re-ran the SED fits.
The resulting AGN disk luminosity was relatively insensitive to these changes, with a mean offset of $-2$\% (+6\%) for lower (higher) uncertainties.
SFR and $M_*$ are more sensitive, with mean SFR offsets of $-15$\% (+33\%) and mean $M_*$ offsets +8\% ($-12$\%) for the same changes.
This bias in output parameter is still within the quoted uncertainties, and we consider a detailed investigation to be better suited to SED-fitting analyses 
incorporating more photometric bands and/or spectroscopic redshifts.
}

\begin{figure}
\centering
\includegraphics[width=\columnwidth]{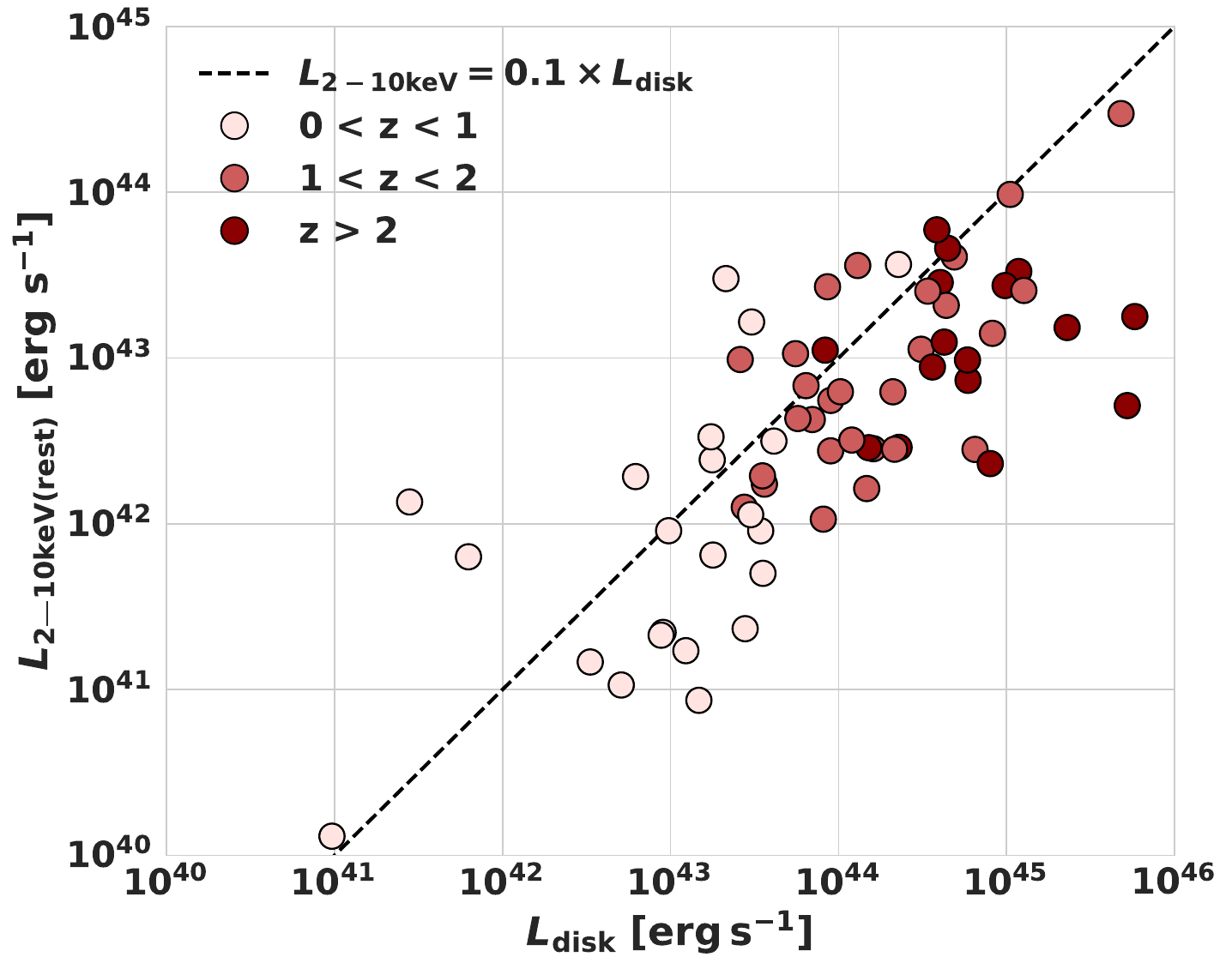}
\caption{The rest-frame \SIrange{2}{10}{\kilo\electronvolt} luminosity of each source versus the integrated luminosity of the AGN disk estimated by CIGALE. This does not account for any light obscured and re-emitted by the dusty torus. The black dashed line represents the 10:1 ratio between the X-ray luminosity and the disk luminosity.}
\label{fig:AGN_Luminosity}
\end{figure}

With this sample of relatively low-luminosity AGN and a simple SED modeling, we are able to extract physical information about the AGN. Figure \ref{fig:AGN_Luminosity} shows the X-ray luminosity of the source as a function of the disc luminosity estimated by CIGALE. We find that our estimates tend to align with the expected ratio of between 5:1 to 20:1 of bolometric to X-ray flux (considering $L_{\rm disk}$ as $L_{\rm bol}$ found for the X-ray selected AGN sample of \citet{Lusso2012}. We do not include the contribution of light from the accretion disk absorbed and re-emitted by an obscuring dusty torus as the constraints on the torus contribution are weak because (1) the uncertainties on the AGN contribution to the light at longer wavelengths \added{are greater}, (2) the limited wavelength coverage of the photometry \added{does not extend} to the rest-frame mid-IR \added{where the torus contribution peaks, and (3) the residuals from the PSF AGN model fits often show positive excesses for the longest wavelength photometry, suggesting an additional contribution to the photometry (likely from the galaxy) in those bands (see Fig.~\ref{fig:Ex_SED_Fit}).  In contrast, (1) the smaller PSF at shorter wavelengths, (2) the tendency of AGN to be bluer than galaxies, and (3) the robust model fits (with symmetric residuals) to the disk component that dominates in the rest-frame optical adds confidence to the $L_{\rm disk}$ values.} The consistency with expectations \added{for the sample compared to the literature lends further credence to our claim that our methodology has successfully extracted the AGN light contribution from the imaging data.}

\section{Conclusions} \label{sec:conclusion2}

In this paper, we develop and evaluate parametric and nonparametric techniques used in tandem to disentangle the light from an AGN from that of its host galaxy for multi-band, space-based imaging.  We investigated 96 sources in the CEERS survey with redshifts between $0.1<z<4$ imaged by up to 13 bands of HST and JWST data. We were able to successfully construct empirical PSF models for all 13 filters using \textsc{photutils}. Using the PSFs, we apply the surface brightness-fitting software \textsc{AstroPhot} to simultaneously model galaxies across every filter.

Comparing the results of the JWST/NIRCam F115W component of the  PSF+S\'ersic joint models with the results of \citet{Dewsnap2025}, we find that the joint model tends to result in slightly fainter AGN components with higher S\'ersic indices than the single filter models. Additionally, the uncertainties of the best fit parameters are heavily underestimated. However, the JWST images are fit reasonably well with our models, and the \added{unresolved light -- predominantly from the AGN -- } within the highly structured JWST PSF can be well-determined.  Parametric fits of the images in HST filters were consistently poor, with a S\'ersic index within $n\sim1$ regardless of wavelength, galaxy morphology, size, or brightness. \added{As there is no overlapping JWST imaging at shorter wavelengths to directly compare with HST, it is still possible that the vast majority of AGN host galaxies are predominantly disk-dominated at short wavelengths leading to these low values of $n$, but we consider this unlikely.  Instead, we conclude that HST imaging is not able to adequately model spheroidal S\'ersic profiles with an unresolved source in the galaxy nucleus.   In addition to the higher spatial resolution of JWST, the additional sensitivity of the larger telescope means that low surface brightness features at larger radii can be detected and incorporated into the model fits; the light profile beyond the effective radius is an important constraint for determining $n$.} While the HST/WFC3 filters are largely superseded by JWST/NIRCam, being able to accurately measure the flux in the HST/ACS bands would provide useful information about the rest-frame UV for the majority of these sources, resulting in higher quality SED models and physical parameter estimates.

After fitting the S\'ersic+PSF models, we performed an AGN subtraction for each source using the constraints on the PSF model. This resulted in clean images of the host galaxy in each filter. Using \textsc{statmorph} on these images, we found that nonparametric measures typically align with their parametric counterparts from the joint fits. Specifically, the concentration and S\'ersic index follow the relations from \citet{Andrae2011} and the half-light radii and axis ratios strongly agree.  

After obtaining flux measurements via both parametric and nonparametric methods, we used CIGALE to fit SED models to both sets of photometry for the 66 objects in the sample with sufficient data. In general, we find that using the aperture photometry for the galaxies results in better fits as measured by reduced $\chi^2$. In particular, the aperture photometry tends to perform better for sources with clear structure (such as clumps or spiral arms), as the S\'ersic profile is unable to adequately model such complexities.  Therefore, using the parametric fits to determine and subtract off the AGN light is the recommended first step.  The preferred second step (using either parametric or nonparametric measures to characterize the galaxy properties) depends on the specific question to be addressed as well as the morphology of the galaxy (e.g., structured or smooth).  Specifically, nonparametric techniques are recommended to obtain photometry and structural parameters for clumpy galaxies.

Using the best-fit SED models for both the host galaxy and the AGN, we estimate physical properties of the sources including the stellar mass and star formation rate for the galaxies, and the disk luminosity for the AGN (see Tables~\ref{tab:SED_fits_glx} and \ref{tab:SED_fits_agn} for the results). The host galaxies are consistent with the high stellar mass end of a sample of non-AGN galaxies in the SFR vs. $M_*$ space (Fig. \ref{fig:SFR_vs_Stellar_Mass}).  This relatively low luminosity X-ray-selected AGN sample shows a distribution of $L_X$ vs. $L_{disk}$ consistent with the X-ray-selected AGN sample of \citet{Lusso2012}.  We find that the dust contribution of the AGN from the CIGALE fits is overestimated, likely caused by our S\'ersic+PSF fit over-subtracting the AGN component at longer wavelengths due to the lower resolution of the imaging coupled with more highly concentrated galaxy profiles.  Therefore, the $L_{dust}$ values from CIGALE are not reliable in this context.  

We have therefore documented a successful methodology of parametric and nonparametric techniques used in tandem to cleanly separate the light of an AGN from its host galaxy in multi-band JWST imaging.  This separation allows the host galaxy structural properties to be determined without joint-fitting the 2D light distribution in the images.  Furthermore, the photometry in the spectral energy distributions of the galaxy and AGN can be fit independently (rather than jointly), thus reducing the risk of degeneracy increasing the uncertainties in the extracted physical properties.

\begin{acknowledgments}
The authors thank the anonymous referee for a helpful report. We thank J. Kartaltepe and P. Aromal for helpful discussions. We acknowledge the support of the Natural Sciences and Engineering Research Council of Canada (NSERC), [funding reference numbers RGPIN-2024-04039 (P.B.) and RGPIN-2021-04157 (S.C.G)]. Cette recherche a été financée par le Conseil de recherches en sciences naturelles et en génie du Canada (CRSNG), [numéros de référence RGPIN-2024-04039 (P.B.) et RGPIN-2021-04157 (S.C.G)]. C.D. thanks the Ontario Graduate Scholarship (OGS) for their support throughout this project. S.C.G. acknowledges the support of the Canadian Space Agency and a Western Research Leadership Chair Award. We acknowledge the Anishinaabek, Haudenosaunee, and L\=unaap\'eewak Nations, on whose traditional territories our research was produced.

This work is based in part on observations taken by the CANDELS Multi-Cycle Treasury Program with the NASA/ESA Hubble Space Telescope, which is operated by the Association of Universities for Research in Astronomy, Inc., under NASA contract NAS5-26555. This work is based in part on observations made with the NASA/ESA/CSA James Webb Space Telescope. The data were obtained from the Mikulski Archive for Space Telescopes at the Space Telescope Science Institute, which is operated by the Association of Universities for Research in Astronomy, Inc., under NASA contract NAS5-03127 for JWST. These observations are associated with program \#1345. The authors acknowledge the CEERS team led by PI S. Finkelstein for developing their observing program with a zero-exclusive-access period.

\end{acknowledgments}

\facilities{
HST(WFC3, ACS),
JWST(NIRCAM)
}

\software{ 
AstroPhot \citep{Stone2023},
CIGALE \citep{Boquien2019, Yang2022},
photutils \citep{Bradley2022},
SExtractor \citep{Bertin1996},
statmorph \citep{Rodriguez-Gomez2019}
}

\bibliography{references}{}
\bibliographystyle{aasjournalv7}

\end{document}

%% file: TabSED.tex
\begin{longrotatetable}
\movetabledown=10mm
\begin{deluxetable*}{llrrrrrrrrrrrrrr}
\tabletypesize{\scriptsize}
\tablecaption{CIGALE spectral energy distribution fit outputs: galaxy properties
\label{tab:SED_fits_glx}}
\tablehead{
\colhead{Source} & \colhead{Model} & 
\colhead{M*} &  \colhead{unc}  & 
\colhead{SFR} &  \colhead{unc} & 
\colhead{L*} &  \colhead{unc} & 
\colhead{SFH age} &  \colhead{unc} & 
\colhead{Mdust} &  \colhead{unc} & 
\colhead{Ldust} &  \colhead{unc} & 
\colhead{Mgas} & \colhead{unc} \\
\colhead{} & \colhead{} &
\colhead{[M$_\sun$]} & \colhead{[M$_\sun$]} & 
\colhead{[M$_\sun$~yr$^{-1}$]} & \colhead{[M$_\sun$~yr$^{-1}$]} & 
\colhead{[\si{\erg\per\second}]} & \colhead{[\si{\erg\per\second}]} & 
\colhead{[Myr]} & \colhead{[Myr]} & 
\colhead{[M$_\sun$]} & \colhead{[M$_\sun$]} & 
\colhead{[\si{\erg\per\second}]} & \colhead{[\si{\erg\per\second}]} & 
\colhead{[M$_\sun$]} & \colhead{[M$_\sun$]}  
} 
\startdata
AEGIS\_349 & P & 1.65e10 & 1.53e10 & 39.00 & 28.13 & 1.47e45 & 8.14e44 & 3.49e03 & 2.13e03 & 1.16e08 & 1.41e08 & 9.55e44 & 6.71e44 & 1.13e10 & 1.31e10 \\
AEGIS\_366 & P & 6.10e10 & 2.48e10 & 43.59 & 52.81 & 2.15e45 & 1.65e45 & 2.92e03 & 1.05e03 & 1.51e08 & 2.34e08 & 1.52e45 & 1.53e45 & 4.21e10 & 2.18e10 \\
AEGIS\_368 & P & 2.54e10 & 2.22e10 & 107.37 & 61.16 & 3.85e45 & 1.80e45 & 3.02e03 & 1.87e03 & 3.04e08 & 3.37e08 & 3.05e45 & 1.65e45 & 1.62e10 & 1.87e10 \\
AEGIS\_377\_1 & P & 7.42e10 & 1.43e10 & 52.91 & 25.32 & 3.07e45 & 9.98e44 & 1.87e03 & 3.42e02 & 2.15e08 & 2.20e08 & 2.15e45 & 9.75e44 & 4.45e10 & 1.14e10 \\
AEGIS\_378 & P & 3.41e10 & 3.11e10 & 134.73 & 80.57 & 4.82e45 & 2.40e45 & 3.16e03 & 1.88e03 & 3.84e08 & 4.32e08 & 3.73e45 & 2.17e45 & 2.25e10 & 2.66e10 \\
AEGIS\_379 & P & 2.62e10 & 1.69e10 & 16.14 & 22.71 & 7.00e44 & 6.90e44 & 4.61e03 & 1.96e03 & 6.13e07 & 1.13e08 & 4.46e44 & 5.81e44 & 2.08e10 & 1.55e10 \\
AEGIS\_381\_1 & S & 1.35e11 & 1.69e10 & 2.16 & 2.12 & 8.34e44 & 1.01e44 & 3.86e03 & 3.44e02 & 3.71e07 & 3.34e07 & 3.70e44 & 9.62e43 & 1.07e11 & 1.53e10 \\
AEGIS\_407\_1 & P & 1.40e09 & 1.45e09 & 6.19 & 3.19 & 2.20e44 & 9.22e43 & 3.50e03 & 2.17e03 & 5.19e06 & 9.67e06 & 4.23e43 & 5.25e43 & 9.22e08 & 1.23e09 \\
AEGIS\_409\_1 & S & 8.87e10 & 5.50e10 & 77.47 & 93.41 & 3.32e45 & 2.84e45 & 3.81e03 & 1.73e03 & 2.51e08 & 4.04e08 & 2.47e45 & 2.59e45 & 6.63e10 & 4.94e10 \\
AEGIS\_420 & P & 8.36e08 & 8.26e08 & 2.35 & 1.58 & 8.56e43 & 4.65e43 & 3.62e03 & 2.16e03 & 7.39e05 & 2.22e06 & 5.14e42 & 1.13e43 & 5.88e08 & 7.24e08 \\
AEGIS\_423 & P & 8.70e10 & 6.17e10 & 86.51 & 97.58 & 3.52e45 & 2.94e45 & 4.25e03 & 2.04e03 & 2.75e08 & 4.28e08 & 2.67e45 & 2.68e45 & 6.66e10 & 5.58e10 \\
AEGIS\_431 & P & 1.10e10 & 6.11e09 & 56.07 & 32.78 & 2.06e45 & 9.80e44 & 1.70e03 & 9.39e02 & 1.59e08 & 1.81e08 & 1.58e45 & 9.39e44 & 5.89e09 & 4.51e09 \\
AEGIS\_432 & P & 5.34e10 & 4.16e10 & 56.11 & 60.92 & 2.23e45 & 1.82e45 & 4.31e03 & 2.04e03 & 1.13e08 & 2.28e08 & 8.31e44 & 1.18e45 & 4.12e10 & 3.74e10 \\
AEGIS\_434\_2 & S & 2.40e10 & 8.37e09 & 52.75 & 42.27 & 2.29e45 & 1.34e45 & 1.41e03 & 6.19e02 & 1.64e08 & 2.19e08 & 1.63e45 & 1.31e45 & 1.31e10 & 6.45e09 \\
AEGIS\_436\_1 & P & 4.48e10 & 2.52e10 & 86.74 & 77.44 & 3.50e45 & 2.35e45 & 2.44e03 & 1.21e03 & 2.70e08 & 3.64e08 & 2.69e45 & 2.19e45 & 2.88e10 & 2.09e10 \\
AEGIS\_450\_1 & P & 1.73e10 & 7.40e09 & 15.94 & 18.07 & 7.45e44 & 5.60e44 & 2.76e03 & 1.10e03 & 5.36e07 & 8.11e07 & 5.37e44 & 5.21e44 & 1.17e10 & 6.42e09 \\
AEGIS\_455 & S & 9.76e10 & 7.34e10 & 158.87 & 148.51 & 6.19e45 & 4.48e45 & 3.45e03 & 1.84e03 & 5.00e08 & 6.97e08 & 4.74e45 & 4.09e45 & 7.03e10 & 6.47e10 \\
AEGIS\_456 & S & 6.48e10 & 3.58e10 & 63.35 & 71.62 & 2.75e45 & 2.14e45 & 3.38e03 & 1.35e03 & 2.04e08 & 3.04e08 & 2.07e45 & 1.97e45 & 4.62e10 & 3.09e10 \\
AEGIS\_457\_1 & P & 1.29e10 & 7.00e09 & 85.11 & 40.55 & 3.07e45 & 1.20e45 & 1.61e03 & 9.47e02 & 2.41e08 & 2.52e08 & 2.39e45 & 1.16e45 & 6.56e09 & 5.05e09 \\
AEGIS\_463 & P & 2.33e10 & 1.02e10 & 69.83 & 54.57 & 2.82e45 & 1.67e45 & 1.38e03 & 6.25e02 & 1.98e08 & 2.71e08 & 1.97e45 & 1.63e45 & 1.26e10 & 7.58e09 \\
AEGIS\_470\_1 & S & 8.29e09 & 4.85e09 & 47.43 & 24.99 & 1.72e45 & 7.42e44 & 1.69e03 & 9.46e02 & 1.35e08 & 1.46e08 & 1.34e45 & 7.05e44 & 4.39e09 & 3.57e09 \\
AEGIS\_475 & P & 7.30e10 & 3.10e10 & 386.85 & 206.76 & 1.46e46 & 6.21e45 & 1.17e03 & 6.18e02 & 1.16e09 & 1.25e09 & 1.16e46 & 6.03e45 & 3.54e10 & 2.18e10 \\
AEGIS\_480\_2 & P & 3.64e10 & 1.01e10 & 20.32 & 20.72 & 1.17e45 & 7.45e44 & 2.38e03 & 7.54e02 & 7.21e07 & 1.12e08 & 7.18e44 & 7.20e44 & 2.40e10 & 9.28e09 \\
AEGIS\_481 & P & 1.58e10 & 6.24e09 & 37.42 & 32.20 & 1.57e45 & 1.00e45 & 1.44e03 & 6.16e02 & 1.08e08 & 1.57e08 & 1.07e45 & 9.82e44 & 8.76e09 & 4.75e09 \\
AEGIS\_485 & S & 1.32e10 & 7.35e09 & 75.60 & 40.90 & 2.75e45 & 1.22e45 & 1.67e03 & 9.44e02 & 2.12e08 & 2.34e08 & 2.10e45 & 1.17e45 & 6.95e09 & 5.38e09 \\
AEGIS\_487 & P & 9.63e09 & 7.24e09 & 20.93 & 17.04 & 8.00e44 & 5.09e44 & 2.94e03 & 1.55e03 & 6.36e07 & 8.15e07 & 6.27e44 & 4.68e44 & 6.57e09 & 6.16e09 \\
AEGIS\_488 & P & 2.55e11 & 7.30e10 & 194.52 & 141.67 & 1.05e46 & 4.66e45 & 2.33e03 & 7.21e02 & 7.48e08 & 8.48e08 & 7.48e45 & 4.44e45 & 1.60e11 & 6.08e10 \\
AEGIS\_495 & P & 5.18e10 & 1.09e10 & 124.51 & 38.80 & 5.77e45 & 1.29e45 & 8.97e02 & 2.03e02 & 4.47e08 & 4.07e08 & 4.46e45 & 1.26e45 & 2.46e10 & 6.71e09 \\
AEGIS\_497 & S & 3.56e10 & 5.93e09 & 18.90 & 10.92 & 1.19e45 & 4.36e44 & 1.91e03 & 2.84e02 & 7.81e07 & 8.58e07 & 7.78e44 & 4.28e44 & 2.20e10 & 4.82e09 \\
AEGIS\_503\_1 & P & 1.06e10 & 6.05e09 & 24.39 & 20.98 & 9.64e44 & 6.38e44 & 2.33e03 & 1.23e03 & 7.41e07 & 1.00e08 & 7.33e44 & 5.98e44 & 6.70e09 & 4.95e09 \\
AEGIS\_509\_3 & S & 7.96e10 & 1.98e10 & 81.30 & 37.54 & 4.31e45 & 1.31e45 & 1.72e03 & 4.67e02 & 3.17e08 & 3.12e08 & 3.17e45 & 1.26e45 & 4.59e10 & 1.53e10 \\
AEGIS\_511 & P & 5.04e10 & 1.20e10 & 51.18 & 35.15 & 2.70e45 & 1.30e45 & 1.65e03 & 5.17e02 & 1.81e08 & 2.26e08 & 1.80e45 & 1.29e45 & 2.93e10 & 9.82e09 \\
AEGIS\_512\_1 & S & 1.44e08 & 5.30e07 & 1.58 & 0.17 & 5.33e43 & 4.64e42 & 4.01e03 & 2.08e03 & 2.62e06 & 2.69e06 & 4.51e43 & 4.43e42 & 8.30e07 & 4.55e07 \\
AEGIS\_519\_1 & P & 2.92e10 & 8.47e09 & 19.86 & 18.90 & 1.08e45 & 6.39e44 & 2.36e03 & 7.41e02 & 7.29e07 & 9.98e07 & 7.29e44 & 6.10e44 & 1.89e10 & 7.42e09 \\
AEGIS\_520 & P & 2.03e10 & 5.04e09 & 11.86 & 9.19 & 6.90e44 & 3.54e44 & 2.36e03 & 7.25e02 & 4.48e07 & 5.83e07 & 4.47e44 & 3.43e44 & 1.31e10 & 4.63e09 \\
AEGIS\_525\_3 & S & 2.60e10 & 1.65e09 & 1.43 & 0.47 & 2.47e44 & 1.90e43 & 2.99e03 & 7.46e01 & 1.32e07 & 1.12e07 & 1.31e44 & 1.79e43 & 1.92e10 & 1.33e09 \\
AEGIS\_532 & P & 1.66e11 & 2.97e10 & 69.68 & 31.23 & 4.77e45 & 1.16e45 & 1.99e03 & 1.18e02 & 3.21e08 & 3.05e08 & 3.21e45 & 1.12e45 & 1.05e11 & 2.24e10 \\
AEGIS\_535 & P & 1.53e10 & 1.30e10 & 22.86 & 21.06 & 8.79e44 & 6.25e44 & 4.04e03 & 2.08e03 & 2.81e07 & 6.17e07 & 2.06e44 & 3.17e44 & 1.14e10 & 1.16e10 \\
AEGIS\_537\_1 & P & 8.39e09 & 4.77e09 & 50.01 & 26.07 & 1.81e45 & 7.74e44 & 1.67e03 & 9.48e02 & 1.34e08 & 1.48e08 & 1.33e45 & 7.42e44 & 4.38e09 & 3.46e09 \\
AEGIS\_540\_1 & P & 2.95e10 & 2.13e10 & 26.05 & 31.20 & 1.07e45 & 9.42e44 & 4.33e03 & 2.03e03 & 8.65e07 & 1.49e08 & 6.36e44 & 7.61e44 & 2.28e10 & 1.93e10 \\
AEGIS\_563\_1 & S & 5.67e10 & 7.94e09 & 17.73 & 5.89 & 1.36e45 & 2.29e44 & 2.00e03 & 3.80e01 & 8.88e07 & 7.95e07 & 8.85e44 & 2.23e44 & 3.63e10 & 5.67e09 \\
AEGIS\_565 & P & 1.84e10 & 1.50e10 & 56.46 & 38.44 & 2.09e45 & 1.14e45 & 2.78e03 & 1.57e03 & 1.68e08 & 1.97e08 & 1.65e45 & 1.05e45 & 1.21e10 & 1.26e10 \\
AEGIS\_570 & S & 3.11e10 & 2.75e10 & 50.91 & 44.41 & 1.96e45 & 1.31e45 & 3.87e03 & 2.12e03 & 1.59e08 & 2.17e08 & 1.24e45 & 1.05e45 & 2.29e10 & 2.43e10 \\
AEGIS\_589\_1 & P & 3.95e10 & 9.00e09 & 33.77 & 21.00 & 1.86e45 & 7.73e44 & 1.74e03 & 4.57e02 & 1.28e08 & 1.46e08 & 1.28e45 & 7.57e44 & 2.34e10 & 7.33e09 \\
AEGIS\_593\_2 & S & 1.24e09 & 6.33e08 & 0.37 & 0.70 & 1.82e43 & 2.09e43 & 5.29e03 & 1.59e03 & 9.85e05 & 1.98e06 & 1.30e43 & 1.88e43 & 1.03e09 & 5.83e08 \\
AEGIS\_599\_3 & P & 1.72e10 & 1.50e10 & 48.22 & 34.94 & 1.77e45 & 1.04e45 & 3.56e03 & 2.16e03 & 1.40e08 & 1.76e08 & 1.26e45 & 9.38e44 & 1.18e10 & 1.30e10 \\
AEGIS\_600\_1 & P & 3.53e10 & 2.80e10 & 58.32 & 54.47 & 2.25e45 & 1.64e45 & 3.86e03 & 2.13e03 & 1.66e08 & 2.53e08 & 1.42e45 & 1.43e45 & 2.57e10 & 2.47e10 \\
AEGIS\_616\_1 & P & 1.72e10 & 1.18e10 & 22.57 & 23.83 & 9.15e44 & 7.24e44 & 3.48e03 & 1.83e03 & 6.66e07 & 1.09e08 & 5.85e44 & 6.45e44 & 1.24e10 & 1.04e10 \\
AEGIS\_628\_1 & P & 1.70e10 & 1.32e10 & 53.25 & 36.75 & 1.98e45 & 1.09e45 & 2.68e03 & 1.57e03 & 1.56e08 & 1.87e08 & 1.54e45 & 1.01e45 & 1.10e10 & 1.10e10 \\
AEGIS\_630\_3\_A & P & 1.32e11 & 1.66e10 & 39.93 & 12.36 & 3.11e45 & 4.83e44 & 2.00e03 & 3.33e01 & 2.02e08 & 1.79e08 & 2.02e45 & 4.70e44 & 8.46e10 & 1.18e10 \\
AEGIS\_634 & P & 2.92e10 & 2.41e10 & 67.47 & 53.33 & 2.53e45 & 1.60e45 & 3.28e03 & 1.87e03 & 2.10e08 & 2.66e08 & 1.97e45 & 1.45e45 & 2.04e10 & 2.09e10 \\
AEGIS\_635 & S & 4.78e09 & 3.41e09 & 10.23 & 8.61 & 3.96e44 & 2.59e44 & 2.80e03 & 1.56e03 & 2.95e07 & 4.01e07 & 2.85e44 & 2.36e44 & 3.21e09 & 2.87e09 \\
AEGIS\_642 & S & 3.59e10 & 2.58e10 & 138.65 & 87.24 & 5.13e45 & 2.59e45 & 2.19e03 & 1.27e03 & 4.05e08 & 4.64e08 & 3.97e45 & 2.41e45 & 2.16e10 & 2.06e10 \\
AEGIS\_653\_1 & S & 3.12e10 & 2.03e10 & 81.43 & 63.06 & 3.13e45 & 1.89e45 & 2.39e03 & 1.24e03 & 2.48e08 & 3.10e08 & 2.47e45 & 1.75e45 & 1.98e10 & 1.67e10 \\
AEGIS\_657\_4 & P & 6.65e09 & 6.27e09 & 20.41 & 13.50 & 7.40e44 & 4.00e44 & 3.63e03 & 2.16e03 & 5.97e07 & 6.99e07 & 5.84e44 & 3.64e44 & 4.62e09 & 5.46e09 \\
AEGIS\_658\_1 & P & 4.18e10 & 2.55e10 & 104.72 & 79.81 & 4.07e45 & 2.36e45 & 2.44e03 & 1.20e03 & 3.20e08 & 3.88e08 & 3.24e45 & 2.18e45 & 2.62e10 & 2.04e10 \\
AEGIS\_660 & P & 2.24e10 & 1.44e10 & 29.52 & 29.99 & 1.20e45 & 8.90e44 & 3.33e03 & 1.39e03 & 9.02e07 & 1.28e08 & 9.31e44 & 8.15e44 & 1.59e10 & 1.22e10 \\
AEGIS\_664\_1 & P & 3.61e10 & 2.95e10 & 125.48 & 80.66 & 4.61e45 & 2.39e45 & 2.67e03 & 1.58e03 & 3.64e08 & 4.21e08 & 3.56e45 & 2.19e45 & 2.31e10 & 2.45e10 \\
AEGIS\_665\_3\_A & P & 8.63e10 & 2.85e10 & 28.03 & 45.05 & 1.77e45 & 1.36e45 & 2.73e03 & 5.34e02 & 1.18e08 & 1.86e08 & 1.19e45 & 1.24e45 & 6.10e10 & 2.36e10 \\
AEGIS\_665\_3\_B & P & 3.85e10 & 7.23e09 & 7.85 & 6.72 & 6.55e44 & 2.77e44 & 2.74e03 & 4.48e02 & 3.81e07 & 4.69e07 & 3.79e44 & 2.65e44 & 2.70e10 & 6.66e09 \\
AEGIS\_699\_4 & P & 1.15e10 & 6.63e09 & 8.32 & 10.50 & 3.72e44 & 3.12e44 & 4.02e03 & 1.57e03 & 2.54e07 & 3.97e07 & 2.79e44 & 2.85e44 & 8.61e09 & 5.87e09 \\
AEGIS\_704\_1 & P & 4.00e10 & 2.29e10 & 130.74 & 89.42 & 5.04e45 & 2.61e45 & 2.11e03 & 1.21e03 & 2.73e08 & 3.86e08 & 2.63e45 & 2.33e45 & 2.30e10 & 1.70e10 \\
AEGIS\_712 & P & 2.39e11 & 8.35e10 & 144.95 & 159.86 & 7.73e45 & 5.02e45 & 2.95e03 & 9.82e02 & 5.37e08 & 7.48e08 & 5.41e45 & 4.67e45 & 1.63e11 & 7.35e10 \\
AEGIS\_721\_1 & S & 3.15e10 & 3.09e10 & 124.30 & 71.59 & 4.43e45 & 2.12e45 & 3.26e03 & 1.88e03 & 3.45e08 & 3.84e08 & 3.41e45 & 1.90e45 & 2.11e10 & 2.64e10 \\
AEGIS\_726 & S & 1.35e10 & 7.01e09 & 22.25 & 22.28 & 9.19e44 & 6.85e44 & 2.48e03 & 1.18e03 & 6.86e07 & 1.02e08 & 6.80e44 & 6.41e44 & 8.80e09 & 5.87e09 \\
AEGIS\_742\_1 & S & 7.61e10 & 4.62e10 & 322.97 & 185.63 & 1.21e46 & 5.40e45 & 1.79e03 & 9.45e02 & 9.79e08 & 1.05e09 & 9.81e45 & 4.99e45 & 4.23e10 & 3.43e10 
\enddata
\end{deluxetable*}
\end{longrotatetable}

\startlongtable
\begin{deluxetable*}{llrrrrr}
\tablecaption{CIGALE spectral energy distribution fit outputs: AGN properties
\label{tab:SED_fits_agn}}
\tablehead{
\colhead{AEGISXD} & \colhead{Source} &  \colhead{$z$} & 
\colhead{Ldisk} &  \colhead{Ldisk unc} & 
\colhead{Lum} & \colhead{Lum unc}  
}
\startdata
J141857.24+524414.0 & AEGIS\_349 & 0.30 & 3.33e42 & 2.28e41 & 1.37e43 & 2.16e42 \\
J141855.75+524507.1 & AEGIS\_366 & 1.42 & 8.97e43 & 6.80e43 & 2.02e46 & 3.39e46 \\
J141901.47+524509.5 & AEGIS\_368 & 0.84 & 2.27e44 & 1.14e43 & 3.72e44 & 2.07e43 \\
J141900.44+524542.8 & AEGIS\_377\_1 & 2.71 & 1.18e45 & 3.56e45 & 5.73e46 & 4.56e47 \\
J141910.74+524543.5 & AEGIS\_378 & 0.96 & 1.77e43 & 1.29e42 & 7.70e43 & 1.84e44 \\
J141912.01+524547.0 & AEGIS\_379 & 0.46 & 5.08e42 & 2.86e41 & 2.10e43 & 2.83e42 \\
J141907.64+524558.2 & AEGIS\_381\_1 & 1.56 & 6.49e44 & 3.85e44 & 1.17e47 & 9.79e46 \\
J141916.88+524656.7 & AEGIS\_407\_1 & 0.45 & 9.02e42 & 5.15e41 & 1.11e43 & 5.55e41 \\
J141910.69+524702.4 & AEGIS\_409\_1 & 0.82 & 2.14e43 & 1.88e42 & 9.74e43 & 9.05e43 \\
J141902.50+524720.4 & AEGIS\_420 & 0.17 & 9.65e40 & 4.82e39 & 1.15e41 & 5.74e39 \\
J141923.47+524727.3 & AEGIS\_423 & 0.74 & 4.12e43 & 3.27e42 & 1.75e44 & 2.61e43 \\
J141910.70+524751.4 & AEGIS\_431 & 1.91 & 4.88e44 & 2.38e44 & 1.85e46 & 1.20e46 \\
J141926.12+524757.7 & AEGIS\_432 & 0.44 & 2.78e43 & 1.48e42 & 3.39e43 & 1.69e42 \\
J141905.40+524759.8 & AEGIS\_434\_2 & 2.59 & 5.90e44 & 6.20e44 & 4.68e46 & 2.06e47 \\
J141909.20+524807.3 & AEGIS\_436\_1 & 1.50 & 1.61e44 & 2.05e43 & 8.97e44 & 4.38e45 \\
J141927.23+524856.2 & AEGIS\_450\_1 & 1.56 & 8.59e43 & 3.25e43 & 1.13e45 & 8.19e45 \\
J141914.08+524909.1 & AEGIS\_455 & 0.82 & 1.74e43 & 1.29e42 & 6.60e43 & 1.01e43 \\
J141930.53+524910.0 & AEGIS\_456 & 1.20 & 8.11e43 & 5.06e42 & 3.15e44 & 5.23e44 \\
J141913.12+524911.3 & AEGIS\_457\_1 & 2.04 & 8.00e44 & 6.53e44 & 7.47e46 & 5.89e46 \\
J141906.44+524933.2 & AEGIS\_463 & 2.27 & 5.85e44 & 4.93e44 & 3.17e46 & 4.14e46 \\
J141859.87+524946.1 & AEGIS\_470\_1 & 1.78 & 3.10e44 & 2.91e44 & 1.46e46 & 1.99e46 \\
J141943.87+524958.6 & AEGIS\_475 & 3.19 & 5.80e45 & 6.16e45 & 3.39e47 & 2.85e48 \\
J141903.38+525007.5 & AEGIS\_480\_2 & 2.11 & 1.51e44 & 2.48e43 & 3.60e45 & 1.69e46 \\
J141908.56+525009.8 & AEGIS\_481 & 2.30 & 9.80e44 & 7.69e44 & 5.14e46 & 2.93e46 \\
J141932.60+525023.2 & AEGIS\_485 & 1.95 & 8.24e44 & 7.20e44 & 5.60e46 & 5.62e46 \\
J141908.49+525024.8 & AEGIS\_487 & 1.12 & 3.62e43 & 2.56e42 & 2.03e44 & 2.48e43 \\
J141947.51+525026.4 & AEGIS\_488 & 2.00 & 1.27e45 & 1.11e45 & 2.56e47 & 3.58e47 \\
J141929.07+525042.4 & AEGIS\_495 & 3.42 & 5.23e45 & 1.06e46 & 4.54e47 & 5.87e47 \\
J141941.09+525044.8 & AEGIS\_497 & 2.53 & 4.25e44 & 4.63e44 & 4.81e46 & 2.61e47 \\
J141937.98+525057.3 & AEGIS\_503\_1 & 1.55 & 6.97e43 & 2.99e43 & 6.26e44 & 3.59e45 \\
J141935.70+525108.7 & AEGIS\_509\_3 & 2.21 & 4.03e44 & 4.54e44 & 4.16e46 & 7.48e46 \\
J141934.91+525123.6 & AEGIS\_511 & 2.75 & 2.29e45 & 1.82e45 & 9.54e46 & 2.62e47 \\
J141919.38+525123.7 & AEGIS\_512\_1 & 0.27 & 2.80e41 & 1.09e41 & 2.38e42 & 9.46e41 \\
J141949.30+525134.1 & AEGIS\_519\_1 & 1.93 & 4.37e44 & 5.20e44 & 2.18e46 & 4.74e46 \\
J141921.39+525136.7 & AEGIS\_520 & 2.17 & 2.29e44 & 2.46e44 & 1.89e46 & 2.48e46 \\
J141924.97+525140.8 & AEGIS\_525\_3 & 1.70 & 8.99e43 & 8.75e42 & 4.99e44 & 1.20e45 \\
J141924.16+525157.4 & AEGIS\_532 & 2.32 & 3.62e44 & 4.96e44 & 6.22e46 & 6.15e46 \\
J141929.83+525206.5 & AEGIS\_535 & 0.37 & 1.47e43 & 1.42e42 & 2.03e43 & 1.62e42 \\
J141948.90+525206.9 & AEGIS\_537\_1 & 1.84 & 3.40e44 & 1.70e43 & 5.22e44 & 4.23e44 \\
J141915.67+525215.6 & AEGIS\_540\_1 & 0.45 & 8.79e42 & 1.48e42 & 3.03e43 & 3.76e42 \\
J141937.21+525329.8 & AEGIS\_563\_1 & 2.51 & 4.46e44 & 5.18e44 & 4.22e46 & 1.70e47 \\
J142007.00+525343.9 & AEGIS\_565 & 1.10 & 5.54e43 & 2.78e42 & 2.01e44 & 2.05e44 \\
J142003.54+525404.1 & AEGIS\_570 & 0.37 & 1.23e43 & 8.54e41 & 1.55e43 & 8.30e41 \\
J141926.24+525441.3 & AEGIS\_589\_1 & 2.19 & 3.85e44 & 4.65e44 & 2.14e46 & 4.04e46 \\
J141930.03+525448.4 & AEGIS\_593\_2 & 0.75 & 6.27e41 & 9.06e40 & 3.57e42 & 4.78e42 \\
J141939.61+525507.1 & AEGIS\_599\_3 & 0.75 & 3.04e43 & 3.07e42 & 1.77e44 & 3.53e44 \\
J142010.08+525511.9 & AEGIS\_600\_1 & 0.70 & 1.79e43 & 8.94e41 & 6.47e43 & 7.00e42 \\
J141950.77+525612.6 & AEGIS\_616\_1 & 0.76 & 3.44e43 & 4.05e42 & 5.56e43 & 3.07e42 \\
J141939.18+525643.4 & AEGIS\_628\_1 & 1.18 & 5.72e43 & 2.86e42 & 1.86e44 & 1.83e44 \\
J141946.17+525646.3 & AEGIS\_630\_3\_A & 2.22 & 8.32e43 & 2.03e43 & 7.22e45 & 1.88e46 \\
J142015.79+525655.6 & AEGIS\_634 & 0.77 & 9.74e42 & 4.87e41 & 3.31e43 & 4.09e42 \\
J141939.07+525657.5 & AEGIS\_635 & 1.13 & 2.60e43 & 7.32e42 & 1.87e45 & 7.30e44 \\
J141949.24+525704.3 & AEGIS\_642 & 1.28 & 1.20e44 & 9.45e42 & 5.24e44 & 1.31e45 \\
J142029.18+525726.9 & AEGIS\_653\_1 & 1.35 & 1.47e44 & 8.19e42 & 5.54e44 & 7.08e44 \\
J142027.01+525739.0 & AEGIS\_657\_4 & 0.73 & 3.55e43 & 3.24e42 & 1.70e44 & 3.05e43 \\
J142029.11+525740.8 & AEGIS\_658\_1 & 1.40 & 1.05e45 & 5.25e43 & 1.58e45 & 1.36e44 \\
J141954.47+525744.6 & AEGIS\_660 & 1.06 & 2.75e43 & 2.79e42 & 1.15e44 & 5.64e43 \\
J141943.38+525748.6 & AEGIS\_664\_1 & 1.10 & 6.40e43 & 5.31e42 & 2.72e44 & 1.21e44 \\
J142038.36+525749.3 & AEGIS\_665\_3\_A & 1.65 & 1.02e44 & 1.67e43 & 6.85e44 & 2.35e44 \\
J142038.36+525749.3 & AEGIS\_665\_3\_B & 1.65 & 2.11e44 & 9.59e43 & 1.66e45 & 8.61e45 \\
J142029.65+525914.1 & AEGIS\_699\_4 & 0.90 & 6.19e42 & 7.03e41 & 3.42e43 & 6.27e42 \\
J142033.57+525917.0 & AEGIS\_704\_1 & 1.30 & 3.52e43 & 2.16e42 & 1.43e44 & 3.12e44 \\
J142002.64+525933.4 & AEGIS\_712 & 1.50 & 2.15e44 & 2.06e43 & 1.40e45 & 1.51e44 \\
J142002.06+525950.3 & AEGIS\_721\_1 & 0.91 & 3.00e43 & 3.54e42 & 1.77e44 & 3.56e43 \\
J142010.93+525958.4 & AEGIS\_726 & 1.58 & 1.30e44 & 6.50e42 & 1.99e44 & 2.64e43 \\
J142005.60+530036.7 & AEGIS\_742\_1 & 1.64 & 4.80e45 & 5.19e44 & 1.09e46 & 3.05e46 \\
\enddata
\tablecomments{All luminosities are in units of \si{\erg\per\second}.}
\end{deluxetable*}